\documentclass[aps,prl,longbibliography,showpacs,twocolumn,superscriptaddress]{revtex4-2}
\usepackage{amsmath,amssymb,amsfonts,bm}
\usepackage{booktabs}
\usepackage{braket}
\usepackage{graphicx}
\usepackage{epstopdf}
\usepackage{dcolumn}
\usepackage{mathrsfs}
\usepackage{bbold}
\usepackage{dsfont}
\usepackage{float}
\usepackage{soul}
\usepackage[colorlinks=true,linkcolor=blue,citecolor=blue, urlcolor=blue,bookmarks=false]{hyperref}
\usepackage{amsbsy}
\usepackage{stmaryrd}
\usepackage{babel}
\usepackage{tgtermes}
\usepackage{multirow}
\usepackage[title,titletoc]{appendix}

\begin{document}

\title{Floquet Spin-Antiferroelectricity in Collinear Antiferromagnets}

\author{Yu-Hao Wei}
\affiliation{Institute for Structure and Function $\&$ Department of Physics $\&$ Chongqing Key Laboratory for Strongly Coupled Physics, Chongqing University, Chongqing 400044, People's Republic of China}
\author{Zheng Qin}
\affiliation{Institute for Structure and Function $\&$ Department of Physics $\&$ Chongqing Key Laboratory for Strongly Coupled Physics, Chongqing University, Chongqing 400044, People's Republic of China}
\author{Shengpu Huang}
\affiliation{Institute for Structure and Function $\&$ Department of Physics $\&$ Chongqing Key Laboratory for Strongly Coupled Physics, Chongqing University, Chongqing 400044, People's Republic of China}
\author{Dong-Hui Xu}
\affiliation{Institute for Structure and Function $\&$ Department of Physics $\&$ Chongqing Key Laboratory for Strongly Coupled Physics, Chongqing University, Chongqing 400044, People's Republic of China}
\affiliation{Center of Quantum materials and devices, Chongqing University, Chongqing 400044, People's Republic of China}
\author{Da-Shuai Ma}
\email{mads@cqu.edu.cn}
\affiliation{Institute for Structure and Function $\&$ Department of Physics $\&$ Chongqing Key Laboratory for Strongly Coupled Physics, Chongqing University, Chongqing 400044, People's Republic of China}
\affiliation{Center of Quantum materials and devices, Chongqing University, Chongqing 400044, People's Republic of China}
\author{Rui Wang}
\email{rcwang@cqu.edu.cn}
\affiliation{Institute for Structure and Function $\&$ Department of Physics $\&$ Chongqing Key Laboratory for Strongly Coupled Physics, Chongqing University, Chongqing 400044, People's Republic of China}
\affiliation{Center of Quantum materials and devices, Chongqing University, Chongqing 400044, People's Republic of China}
\affiliation{State Key Laboratory of Mechanical Transmission for Advanced Equipment, Chongqing University, Chongqing 400044, People's Republic of China}

\begin{abstract}
Multiferroics combining magnetic and polar orders offer opportunities for optical control of spin and electric degrees of freedom. Here, using symmetry analysis and Floquet theory, we establish Floquet spin-antiferroelectricity coexisting with unconventional magnetism in periodically driven collinear antiferromagnets, qualifying it as an unconventional multiferroic. This driven phase supports compensated, spin-resolved in-plane electric polarizations perpendicular to the vertical rotation axis, which are strictly forbidden by crystalline symmetry in equilibrium. Using a tight-binding model, we elucidate how light polarization controls the emergence of polar order and spin responses. Moreover, first-principles-based Floquet calculations predict its realization in monolayer $\text{MnPS}_3$, identifying a realistic two-dimensional antiferromagnetic platform. These findings establish a nonequilibrium route to spin-antiferroelectricity beyond equilibrium symmetry constraints and connect Floquet engineering, unconventional magnetism, and multiferroicity through optical control of spin and polar degrees of freedom.
\end{abstract}

\maketitle

\textit{\textcolor{blue}{Introduction}}---Multiferroics couple magnetic and polar orders, enabling the manipulation of spin-dependent electronic properties via electric degrees of freedom, thereby offering a versatile platform for high-density data storage and energy-efficient spintronic devices~\cite{Eerenstein2006,Cheong2007,Ramesh2007,Matsukura2015,Fiebig2016,Spaldin2019,Manipatruni2019,Fert2024,Wang2003,Kimura2003,Zhao2006,Chu2008,Heron2014}. 
Conventionally, such coupling is realized by combining real-space polar order with ferromagnetism or antiferromagnetism~\cite{Wang2003,Kimura2003,Zhao2006,Chu2008,Heron2014}.  Recent advances in altermagnetism within the framework of spin group theory have expanded the scope of this emerging  paradigm~\cite{Smejkal2022Phase,Smejkal2022Landscape,Krempasky2024,Amin2024,Song2025,BaiSpinSplitting2022,JiangAltermagnet2025,CrSbControl2025,Jungwirth2026,LeeMnTe2024,litvin1974spin,Liu2022,XiaoSpinGroup2024,JiangSpinSpace2024,PhysRevX.14.031038,LiuNature2026}, giving rise to unconventional multiferroics, such as ferroelectric or antiferroelectric altermagnets~\cite{SunSliding2024,SunStacking2024,BezzergaMnSe2025,
UrruBiFeO32025,SunTypeIII2025,ZhuSliding2025,SunSixState2026,WangDual2025,
SunFramework2026,GuoTypeII2026,SunSpinSymmetry2025,Gu2025,Zhu2025,Duan2025,GuoTriferroic2025,
ChenHeterostructure2026,ZhaoInterconversion2026}, as well as spin-antiferroelectrics where the polarization in momentum space is encoded in electronic wave functions~\cite{Wang2026,FuSpinAFE2026}. Unlike their conventional counterparts, these emergent phases exhibit symmetry-locked magnetoelectric coupling, enabling nonvolatile electrical control of spin textures and robust readout via Hall and magneto-optical effects~\cite{Gu2025,SunSpinSymmetry2025,SunFramework2026,GuoTypeII2026,SunSixState2026,WangDual2025}.

Despite these encouraging advancements, symmetry constraints fundamentally dictate the material realization of both conventional and unconventional multiferroics. 
Specifically, a finite spontaneous polarization is permitted only within polar point groups, whereas any symmetry belonging to the nonpolar set $\mathcal{S}=\{\mathcal{I}, \bar{3}, \bar{4}, \bar{6}\}$ strictly forbids it~\cite{Radaelli2007Symmetry,Cheong2018Broken,Shi2016Symmetry}.
In two-dimensional (2D) antiferromagnets, the presence of rotation about the layer normal ($\mathcal{C}^z_{n \ge 2}$)  forbids in-plane  spontaneous polarization order in both real and momentum space, restricting the available material platforms~\cite{Kruse2023TwoDimensional,Qi2021TwoDimensional,Wang2023Towards}.
Thus, realizing  multiferroics with in-plane polarization  in highly symmetric antiferromagnetic systems remains an open challenge.

Floquet engineering provides a framework for dynamically controlling  crystalline symmetries and their electronic states through periodic driving~\cite{Goldman2014,Bukov2015,Eckardt2017,OkaKitamura2019,deLaTorre2021,ZhanPerspective2024}. Its applications include various  periodically driven topological states of matter~\cite{Jotzu2014,Lindner2011,Rudner2013,Hubener2017,YuFloquet2021,YanWang2016,LiuBlackPhosphorus2018,LiuGraphene2019,WangFloquet2013,Rudner2020,McIver2020,Choi2025,Merboldt2025,WangFloquetGap2026}, light-induced superconductivity~\cite{Fausti2011,Mitrano2016,Ning2024,FuTriplet2026,Claassen2019,Gassner2024,Fava2024,LiuMajorana2013,YangDissipative2021,ZhangChiral2021},  optical control of magnetism~\cite{Nova2017,Afanasiev2021,Mentink2015,ClaassenSpin2017,KitamuraChirality2017,Quito2021,Herre2026,Disa2023,Gorg2018,ZhangCrI3Switch2022,WangMoireMag2022}, as well as beyond~\cite{IkedaPolarization2022,ShanNonlinearity2021,NeufeldFloquet2019,KobayashiExciton2023,ExcitonicFloquet2026}. Recent studies show that periodic driving can induce nonrelativistic odd-parity spin splitting in spin-degenerate collinear antiferromagnets~\cite{Huang2026,ZhuFloquet2026,LiFloquet2026,LiuDimer2026,ZhangTriangulene2026}. Its momentum-odd character contrasts with the even-parity spin splitting of altermagnets~\cite{Smejkal2022Phase,Smejkal2022Landscape,Krempasky2024,Amin2024,Song2025,BaiSpinSplitting2022,JiangAltermagnet2025,CrSbControl2025,Jungwirth2026,LeeMnTe2024}. 
These advances raise the question of whether Floquet engineering can introduce entirely new dimensions to unconventional multiferroics through the dynamic selection of symmetry breaking and preservation.

Here, we propose a new class of unconventional multiferroic, termed the Floquet spin-antiferroelectric, which exhibits the coexistence of in-plane electric polarization and distinct unconventional magnetism  within high-symmetry 2D collinear antiferromagnets. We identify the symmetry requirements and microscopic mechanism for Floquet spin-
antiferroelectricity. Specifically, we reveal that linearly polarized light (LPL) induces compensated in-plane polarization while preserving spin degeneracy, and elliptically polarized light (EPL) removes both the rotational constraint and the dynamical symmetry protecting spin degeneracy, allowing compensated polarization and odd-parity spin splitting to coexist. Additionally, the spin-resolved polarization can be controlled by tuning the polarization angle and relative phase of incident light, providing all-optical control of the driven polar state. 
Finally, using monolayer $\text{MnPS}_3$ as a concrete example, our first-principles-based Floquet calculations  confirm the emergence of this Floquet spin-antiferroelectric state and its accessibility in experimental conditions.

\begin{figure}[t]
    \centering
    \includegraphics[width=\linewidth]{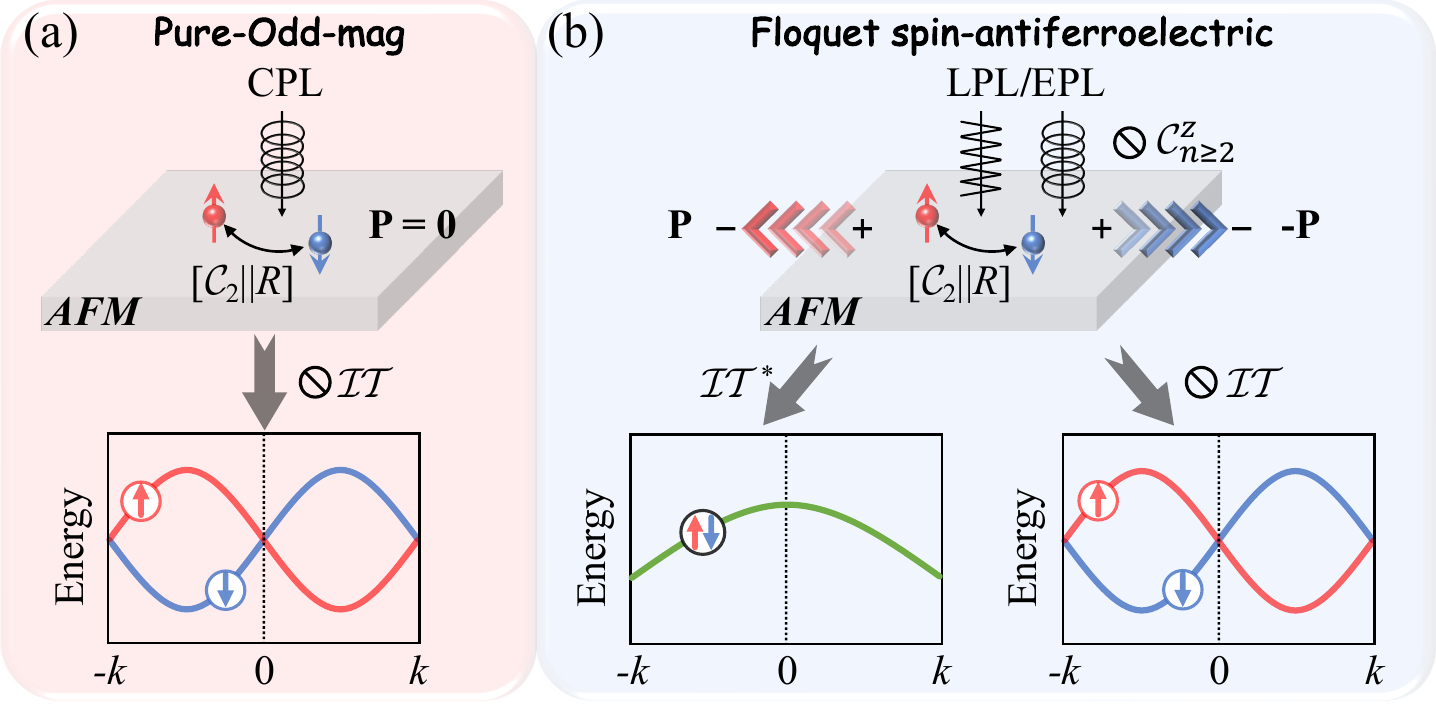}
    \caption{Conceptual diagram of optically engineered Floquet spin-antiferroelectricity.
    (a) CPL breaks the antiunitary symmetry that protects spin degeneracy and allows odd-parity spin splitting. The spin-preserving nonpolar constraint remains intact, so the sector-resolved electric polarization is forbidden.
    (b) Floquet phases controlled by the light polarization state. In the LPL case, linearly polarized light selects a fixed in-plane direction and breaks the spin-preserving nonpolar constraint, allowing a compensated spin-antiferroelectric order. However, a residual dynamical space-time symmetry protects spin degeneracy and suppresses odd-parity spin splitting. In the EPL case, elliptically polarized light breaks both the spin-preserving nonpolar constraint and the dynamical symmetry protecting spin degeneracy, while preserving the spin-reversing compensation symmetry. This realizes a compensated Floquet spin-antiferroelectricity in which odd-parity spin splitting coexists with compensated sector-resolved electric polarizations.}
    \label{fig1}
\end{figure}

\textit{\textcolor{blue}{Symmetry analysis of Floquet spin-antiferroelectricity.}}---According to the modern theory of polarization, the macroscopic electric polarization $\mathbf{P}$ of a band insulator is formulated as a geometric Berry phase~\cite{KingSmith1993,Vanderbilt1993,Resta1994}:
\begin{equation}
\mathbf{P} = -\frac{e}{(2\pi)^D} \sum_{n \in \mathrm{occ}} \int_{\mathrm{BZ}} d^D\boldsymbol{k} \boldsymbol{A}_{n}(\boldsymbol{k}),
\label{eq1}
\end{equation}
where $-e$ is the elementary charge, $D$ is the dimension, and $\boldsymbol{A}_{n}(\boldsymbol{k}) = i \langle u_{n}(\boldsymbol{k}) | \nabla_{\boldsymbol{k}} | u_{n}(\boldsymbol{k}) \rangle$ is the Berry connection. The polarization is well defined modulo the polarization quantum $\mathbf{P}_q^{(D)}=e\mathbf{R}/\Omega_D$, where $\mathbf{R}$ is a lattice vector and $\Omega_D$ is the $D$-dimensional primitive-cell volume. Under point-group operations $R$, the polarization transforms covariantly as a polar vector ($\mathbf{P} \rightarrow R\mathbf{P}$).

Intriguing electric polarization phenomena emerge when the electronic Hilbert space is decoupled into two sectors by a certain symmetry $\mathcal{M}$, as exemplified by the type-II antiferroelectricity in collinear antiferromagnets~\cite{Wang2026}. In the nonrelativistic limit, the Bloch Hamiltonian is block-diagonal in the spin basis, $H_{\mathrm{eq}}(\boldsymbol{k})=h_{\uparrow}(\boldsymbol{k})\oplus h_{\downarrow}(\boldsymbol{k})$, enabling the spin-resolved decomposition of the polarization, $\mathbf{P}=\mathbf{P}_\uparrow + \mathbf{P}_\downarrow$. Here, each sector polarization $\mathbf{P}_\sigma$ ($\sigma = \uparrow, \downarrow$) is evaluated independently within $h_\sigma(\boldsymbol{k})$ using Eq.~\eqref{eq1}. Under this limit, the symmetries of these decoupled sectors are described by the spin group $\mathcal{G} = \mathcal{G}_0 \cup X\mathcal{G}_0$, whose elements are joint operations $[R_s \parallel R_r]$ acting on spin and real space, respectively~\cite{litvin1974spin,Liu2022,XiaoSpinGroup2024,JiangSpinSpace2024,PhysRevX.14.031038,LiuNature2026}. The spin-preserving subgroup $\mathcal{G}_0$ consists of $[S \parallel R_r]$ with $S \in \{E, \mathcal{C}_2\mathcal{T}\} \ltimes SO(2)$, while the spin-reversing coset $X\mathcal{G}_0$ contains $X = [\mathcal{C}_2 \parallel R_r]$.

To permit nonvanishing sector polarizations ($\mathbf{P}_\sigma \neq 0$), the real-space projection of the spin-preserving subgroup, $\mathcal{G}^R_0$, must be a polar point group. A finite spin-resolved polarization emerges only when these nonpolar spatial constraints are relieved through selective symmetry reduction. Under this symmetry-breaking landscape, the relationship between the spin sectors is uniquely governed by the spin-reversing coset $X\mathcal{G}_0$, whose elements $[\mathcal{C}_2 \parallel R_r]$ enforce the parity constraint $P_{\sigma, \alpha} = \eta P_{-\sigma, \alpha}$ (where $\alpha \in \{x, y, z\}$ and $\eta = \pm 1$ is the spatial parity of $R_r$). For $\eta = -1$ (e.g., under the joint operation $[\mathcal{C}_2 \parallel \mathcal{I}]$), this constraint yields $P_{\uparrow, \alpha} = -P_{\downarrow, \alpha} \neq 0$, realizing a compensated spin-antiferroelectric state with vanishing net polarization ($P_{\uparrow, \alpha} + P_{\downarrow, \alpha} = 0$) but a finite AFE order parameter $Q_\alpha = (P_{\uparrow, \alpha} - P_{\downarrow, \alpha})/2 = P_{\uparrow, \alpha}$. Conversely, $\eta = 1$ yields $P_{\uparrow, \alpha} = P_{\downarrow, \alpha} \neq 0$, permitting a net macroscopic polarization.

For widely studied 2D systems with a vertical rotation axis, the presence of $\mathcal{C}_{n \ge 2}^z$ restricts the real-space projection subgroup $\mathcal{G}_0^R$ to a nonpolar point group, thereby forbidding any spontaneous in-plane sector polarization, i.e., $\mathbf{P}_{\sigma, \parallel} = (P_{\sigma, x}, P_{\sigma, y}, 0) = 0$. By applying periodic optical driving, we can dynamically lower the point-group symmetry of $\mathcal{G}_0^R$ to a polar one, releasing the spin-resolved polar vector into the in-plane directions [$\mathbf{Q} = (Q_x, Q_y, 0) \neq 0$], realizing the Floquet spin-antiferroelectricity. To concretely elucidate how to realize Floquet spin-antiferroelectricity, we investigate the impacts of three distinct optical polarization configurations on the electronic polarization and band spin splitting of a spin-degenerate collinear antiferromagnetic (AFM) system. Under circularly polarized light (CPL), as illustrated in Fig.~\ref{fig1}(a), the field breaks space-time inversion $\mathcal{I}\mathcal{T}$, lifting spin degeneracy to yield an odd-parity spin splitting, $\Delta E(\boldsymbol{k}) = -\Delta E(-\boldsymbol{k}) \neq 0$~\cite{Huang2026,ZhuFloquet2026,LiFloquet2026,LiuDimer2026,ZhangTriangulene2026}. 
However, CPL preserves the crystal rotation symmetry $\mathcal{C}_n$ ($n \ge 2$), leaving $\mathcal{G}_0^R$ non-polar and forcing $\mathbf{P}_{\sigma, \parallel} = 0$. 
Under LPL [Fig.~\ref{fig1}(b), left], the uniaxial field oscillation breaks $\mathcal{C}_n$, lowering $\mathcal{G}_0^R$ to a polar group and allowing compensated in-plane sector polarization ($\mathbf{P}_{\uparrow, \parallel} = -\mathbf{P}_{\downarrow, \parallel} \neq 0$) via the spin-reversing symmetry $[\mathcal{C}_2 \parallel R_r]$ (with $\eta = -1$). 
Yet, because LPL preserves a residual dynamical space-time inversion $\mathcal{I}\mathcal{T}^* = \{\mathcal{I}\mathcal{T} | T/2\}$ (where $T$ is the driving period), spin degeneracy remains protected, suppressing spin splitting. 
Under EPL [Fig.~\ref{fig1}(b), right], the combination of in-plane anisotropy and field rotation simultaneously breaks both $\mathcal{C}_n$ and $\mathcal{I}\mathcal{T}^*$. 
Since the spin-reversing symmetry $[\mathcal{C}_2 \parallel R_r]$ remains intact, EPL uniquely permits the coexistence of compensated in-plane polarization ($\mathbf{P}_{\uparrow, \parallel} = -\mathbf{P}_{\downarrow, \parallel} \neq 0$) and odd-parity spin splitting ($\Delta E(\boldsymbol{k}) \neq 0$), realizing a compensated Floquet spin-antiferroelectric phase.

\begin{figure}[t]
    \centering
    \includegraphics[width=\linewidth]{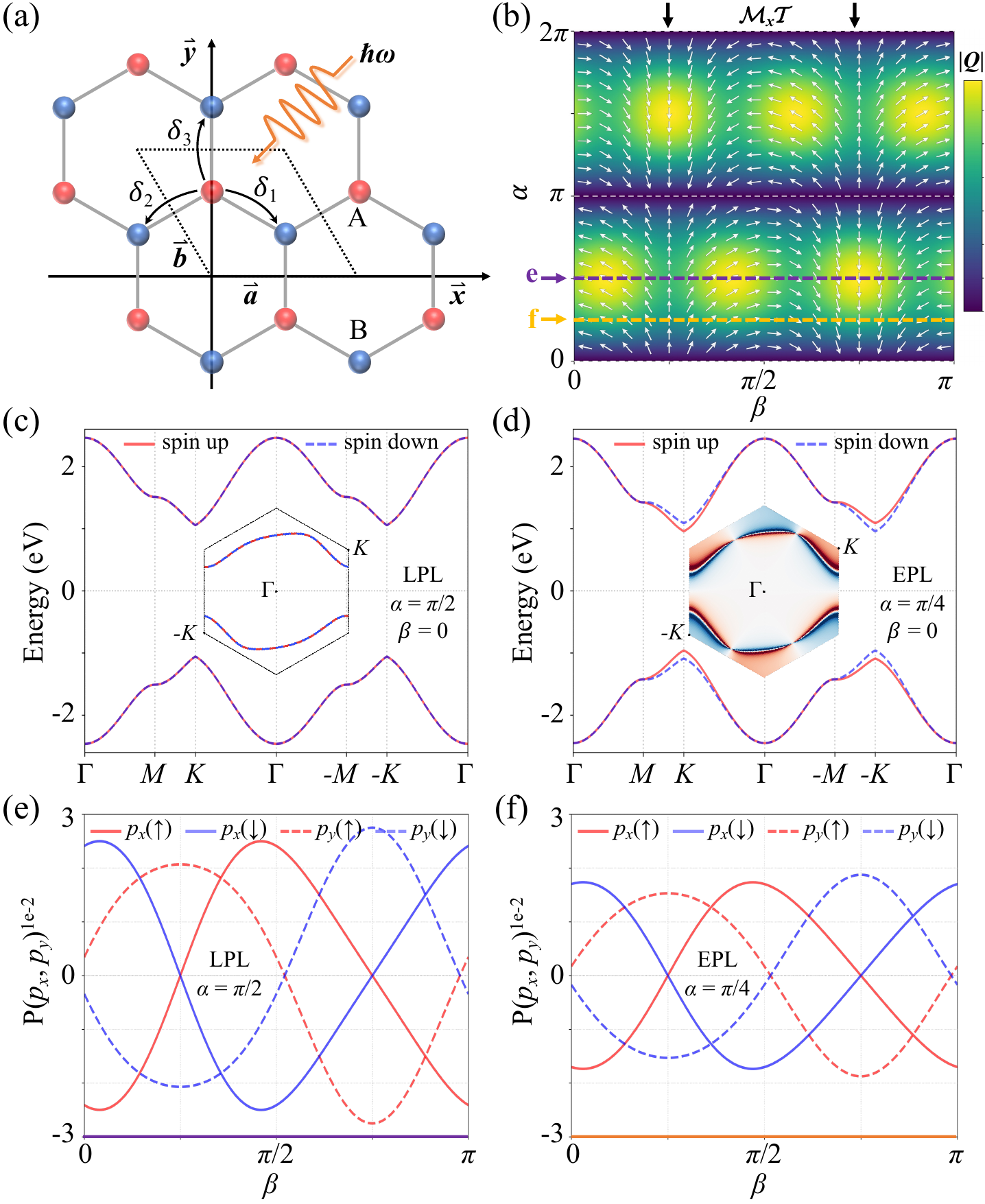}
    \caption{Minimal honeycomb antiferromagnetic tight-binding model for the Floquet spin-antiferroelectricity. 
(a) Honeycomb lattice with collinear AFM order. Red and blue spheres denote A and B sublattices with opposite magnetic moments. 
(b) Evolution of the spin-antiferroelectric order parameter $\mathbf{Q} = (\mathbf{P}_\uparrow - \mathbf{P}_\downarrow)/2$ in the $(\beta, \alpha)$ parameter space. The white arrows show its in-plane direction. Yellow and purple horizontal dashed lines indicate $\alpha = \pi/4$ and $\pi/2$, corresponding to the EPL and LPL cuts plotted in (f) and (e), respectively. Black arrows at $\beta = \pi/4$ and $3\pi/4$ indicate that the system possesses the dynamical $\mathcal{M}_x\mathcal{T}$ symmetry under these parameters. Dashed lines at $\alpha = 0$, $\pi$, and $2\pi$ mark the CPL limits where the remaining threefold rotation enforces $\mathbf{Q} = 0$. 
(c), (d) Spin-resolved Floquet band structures under LPL ($\alpha = \pi/2$) and EPL ($\alpha = \pi/4$), respectively. Insets show spin-resolved isoenergy surfaces at $E = -1.3$~eV. 
(e), (f) Spin-resolved electric polarizations $P_x$ and $P_y$ as functions of the in-plane orientation angle $\beta$ for LPL ($\alpha = \pi/2$) and EPL ($\alpha = \pi/4$). Red and blue curves correspond to the spin-up and spin-down sectors, while solid and dashed curves represent $P_x$ and $P_y$, respectively, under the compensation relation $\mathbf{P}_\uparrow = -\mathbf{P}_\downarrow$. In panels (b)--(f), the model parameters $t=m=1$~eV, the light amplitude $eA_0/\hbar=0.3~\mathrm{\mathring{A}}^{-1}$, and the photon energy $\hbar\omega=20$~eV are adopted.}
    \label{fig2}
\end{figure}

\textit{\textcolor{blue}{Minimal Tight-Binding Model.}}---To elucidate the physical mechanism of this light-induced spin-antiferroelectricity, we construct a minimal tight-binding (TB) model on a 2D honeycomb lattice with collinear AFM order, as schematically illustrated in Fig.~\ref{fig2}(a). The real-space equilibrium Hamiltonian is formulated as
\begin{equation}
H = -t \sum_{\langle ij \rangle} c_i^\dagger c_j + \mathrm{H.c.} + \sum_i m_i \, c_i^\dagger \sigma_z c_i,
\label{eq2}
\end{equation}
where $c_i^\dagger = (c_{i\uparrow}^\dagger, c_{i\downarrow}^\dagger)$ is the electron creation spinor at site $i$ ($i \in A, B$), $t$ is the nearest-neighbor hopping amplitude, and $m_i$ denotes the local exchange field. The collinear N\'eel order is characterized by a staggered exchange field, $m_A = -m_B \equiv m$. By transforming to momentum space with the spinful basis $\Psi^{\dagger}_{\boldsymbol{k}} = (c_{A\boldsymbol{k}\uparrow}^{\dagger}, c_{B\boldsymbol{k}\uparrow}^{\dagger}, c_{A\boldsymbol{k}\downarrow}^{\dagger}, c_{B\boldsymbol{k}\downarrow}^{\dagger})$, the Hamiltonian is expressed as:
\begin{equation}
H(\boldsymbol{k}) = \sigma_0 \otimes 
\begin{pmatrix}
0 & \Delta(\boldsymbol{k}) \\
\Delta^*(\boldsymbol{k}) & 0
\end{pmatrix}
+ m\,\sigma_z \otimes \tau_z ,
\label{eq3}
\end{equation}
where $\sigma_i$ and $\tau_i$ are the Pauli matrices acting on the spin and sublattice spaces, respectively, with $\sigma_0$ being the $2 \times 2$ identity matrix. The off-diagonal hopping term is given by $\Delta(\boldsymbol{k}) = -t \sum_{\ell=1}^{3} e^{i \boldsymbol{k} \cdot \boldsymbol{\delta}_\ell}$, where $\boldsymbol{\delta}_{\ell}$ ($\ell=1,2,3$) are the three nearest-neighbor bond vectors shown in Fig.~\ref{fig2}(a). In the pristine equilibrium state, the system's invariance under the spin-preserving joint operation $[E \parallel \mathcal{C}_{3z}]$ directly forbids any in-plane polar vector, enforcing the in-plane sector polarizations to vanish, i.e., $\mathbf{P}_{\sigma,\parallel} = 0$.

We introduce the optical field through the Peierls substitution and use the vector potential $\boldsymbol{\mathcal A}(t)=A_0[\mathbf e_1\cos(\omega t)+\mathbf e_2\sin(\omega t+\alpha)]$, where the orthogonal polarization axes are defined as $\mathbf e_1=(\cos\beta,\sin\beta,0)$ and $\mathbf e_2=(-\sin\beta,\cos\beta,0)$. Here, $\alpha$ represents the relative optical phase, and $\beta$ denotes the in-plane laser polarization angle. In the off-resonant (high-frequency) regime, where the photon energy is much larger than the electronic bandwidth ($\hbar\omega \gg |t|$) to avoid hybridization between distinct Floquet subbands, the effective Floquet Hamiltonian can be described by the high-frequency expansion~\cite{Goldman2014,Bukov2015,Eckardt2017,OkaKitamura2019,deLaTorre2021,ZhanPerspective2024}:
\begin{equation}
H_{\text{eff}}(\boldsymbol{k}) = H_0(\boldsymbol{k}) + \sum_{n \geq 1} \frac{[H_{-n}(\boldsymbol{k}), H_n(\boldsymbol{k})]}{n\hbar\omega} + \mathcal{O}\left(\frac{1}{(\hbar\omega)^2}\right),
\label{eq:H_eff_general}
\end{equation}
where $H_n(\boldsymbol{k}) = \frac{1}{T}\int_0^T e^{-i n \omega t} H(\boldsymbol{k}, t)\,dt$ is the $n$-th Fourier component of the time-dependent Hamiltonian $H(\boldsymbol{k}, t)$, and $H_0(\boldsymbol{k})$ denotes its time average.

Under CPL with $\alpha=0$ or $\pi$, the threefold rotational symmetry forbids any in-plane sector polarization, as indicated by the white dashed line in Fig.~\ref{fig2}(b).  Therefore, we focus on the spin-resolved polarizations induced by LPL and EPL. For LPL with $\alpha=\pi /2$ and $\beta=0$, the time-averaged hopping amplitudes are anisotropically renormalized as $t_\ell \equiv tJ_0\left(\frac{e A_0}{\hbar} \sqrt{ |\boldsymbol{\delta}_\ell|^2 + 2 (\boldsymbol{e}_1 \cdot \boldsymbol{\delta}_\ell) (\boldsymbol{e}_2 \cdot \boldsymbol{\delta}_\ell) \sin\alpha }\right) = tJ_0\left(\frac{e A_0}{\hbar} |(\boldsymbol{e}_1 + \boldsymbol{e}_2) \cdot \boldsymbol{\delta}_\ell|\right)$, which breaks the $\mathcal C_{3z}$ symmetry. Due to the time-reversal invariance of LPL, the commutator $[H_{n}(\boldsymbol{k}), H_{-n}(\boldsymbol{k})] \equiv 0$ for all $n$, reducing Eq.~\eqref{eq:H_eff_general} strictly to the time-averaged term $H_{\text{eff}}(\boldsymbol{k}) = H_0(\boldsymbol{k})$ (see Supplemental Material (SM)~\cite{SM} for proof). Thus, for the LPL-irradiated AFM system, the spin-resolved effective Hamiltonian reads:
\begin{equation}
H_{\text{eff},\sigma}(\boldsymbol{k}) = \begin{pmatrix} s_\sigma m & \Delta_{\text{eff}}(\boldsymbol{k}) \\ \Delta_{\text{eff}}^{*}(\boldsymbol{k}) & -s_\sigma m \end{pmatrix},
\label{eq:H_LPL}
\end{equation}
where $s_{\uparrow}=+1$ and $s_{\downarrow}=-1$ denote the spin indices, and $\Delta_{\text{eff}}(\boldsymbol{k}) = -\sum_{\ell=1}^3 t_{\ell} e^{i \boldsymbol{k} \cdot \boldsymbol{\delta}_\ell}$. Consequently, both spin channels share the identical energy eigenvalues $E_{\sigma}^{\pm}(\boldsymbol{k}) = \pm \sqrt{(s_\sigma m )^2 + |\Delta_{\text{eff}}(\boldsymbol{k})|^2}$, demonstrating that LPL does not induce any band spin splitting, as shown in Fig.~\ref{fig2}(c).

By parameterizing $\Delta_{\text{eff}}(\boldsymbol k)=|\Delta_{\text{eff}}(\boldsymbol k)|e^{i\theta(\boldsymbol k)}$, the spin-resolved electronic polarization of the occupied band reads,
\begin{equation}
\mathbf P_\sigma=\frac{e}{2(2\pi)^2}\int_{\mathrm{BZ}}d^2\boldsymbol k\left(1-\frac{s_\sigma m}{|E_{\sigma}^{\pm}(\boldsymbol k)|}\right)\nabla_{\boldsymbol k}\theta(\boldsymbol k).
\label{eq:P_analytic}
\end{equation}
In Fig.~\ref{fig2}(e), we present the numerical results of $\mathbf{P}_\sigma$. We find that $\mathbf{P}_\sigma$ is no longer zero and varies periodically with $\beta$. More interestingly, by summing the polarizations of both sectors, one has the total polarization:
\begin{equation}
\mathbf P_{\mathrm{total}}=\mathbf P_\uparrow+\mathbf P_\downarrow
=\frac{e}{(2\pi)^2}\int_{\mathrm{BZ}}d^2\boldsymbol k\,\nabla_{\boldsymbol k}\theta(\boldsymbol k)
=0,
\label{eq:P_total}
\end{equation}
which is guaranteed by the periodic boundary conditions of the Brillouin zone. Consequently, Eq.~\eqref{eq:P_total} establishes that the system maintains a vanishing net polarization $\mathbf{P}_{\mathrm{total}} = 0$ but develops a finite in-plane spin-antiferroelectric order parameter $\mathbf{Q}_{\parallel} \neq 0$ once the threefold rotation symmetry $\mathcal{C}_{3z}$ is broken. This demonstrates the emergence of a light-induced compensated spin-antiferroelectric (type-II AFE) state under LPL, establishing that optical anisotropy can be utilized to generate compensated polarization order.

Next, we discuss the scenario of EPL. The broken time-reversal symmetry of the EPL field generates a real-valued, odd-parity first-order dynamical mass $W(\boldsymbol{k}) \equiv D_1(\boldsymbol{k})/(\hbar\omega) = -W(-\boldsymbol{k})$, modifying the effective Hamiltonian to:
\begin{equation}
H_{\text{eff},\sigma}(\boldsymbol{k}) = \begin{pmatrix} s_\sigma m & \Delta_{\text{eff}}(\boldsymbol{k}) \\ \Delta_{\text{eff}}^{*}(\boldsymbol{k}) & -s_\sigma m \end{pmatrix} + W(\boldsymbol{k}) \tau_z.
\label{eq:Heff_EPL}
\end{equation}
Physically, this momentum-odd mass $W(\boldsymbol{k})\tau_z$ lifts the spin degeneracy, yielding an odd-parity spin splitting $\Delta E(\boldsymbol{k}) = |E_{\uparrow}^{\pm}(\boldsymbol{k})| - |E_{\downarrow}^{\pm}(\boldsymbol{k})| = -\Delta E(-\boldsymbol{k})$, where $E_{\sigma}^{\pm}(\boldsymbol{k}) = \pm  \sqrt{(s_\sigma m + W(\boldsymbol{k}))^2 + |\Delta_{\text{eff}}(\boldsymbol{k})|^2}$, as shown in Fig.~\ref{fig2}(d) for the representative case of $\alpha = \pi/4$. Concurrently, the spin-resolved polarizations of the valence bands are obtained as:
\begin{equation}
\mathbf{P}_\sigma = \frac{e}{2(2\pi)^2}\int_{\mathrm{BZ}} d^2\boldsymbol{k} \left(1 - \frac{s_\sigma m + W(\boldsymbol{k})}{|E_{\sigma}^{\pm}(\boldsymbol{k})|}\right) \nabla_{\boldsymbol{k}}\theta(\boldsymbol{k}).
\label{eq:P_EPL_final}
\end{equation}
Notably, although the individual sector polarizations $\mathbf{P}_\sigma$ are modified by $W(\boldsymbol{k})$, the spin-reversing compensation symmetry $[\mathcal{C}_2 \parallel \mathcal{I}]$ maps the valence states of spin-up at $\boldsymbol{k}$ to spin-down at $-\boldsymbol{k}$. This symmetry constraint physically guarantees a strictly compensated in-plane spin-antiferroelectric state ($\mathbf{P}_{\uparrow} = -\mathbf{P}_{\downarrow} \neq 0$), as exemplified in Fig.~\ref{fig2}(f) under $\alpha = \pi/4$.

\textit{\textcolor{blue}{Optical Tunability of the Spin-Antiferroelectric Order.}}---In Fig.~\ref{fig2}(b), we plot the phase diagram of the AFE order parameter $\mathbf{Q} = (\mathbf{P}_\uparrow - \mathbf{P}_\downarrow)/2$ in the parameter space $(\beta, \alpha)$. Through this phase diagram, we find that $\mathbf{Q}$ is subject to two distinct symmetry constraints. First, because shifting $(\alpha, \beta) \rightarrow (\alpha+\pi, \beta+\pi/2)$ represents a spin-preserving antiunitary time-reversal $\boldsymbol{\mathcal{A}}(t) \rightarrow -\boldsymbol{\mathcal{A}}(-t)$, we obtain the parameter-space translation relation $\mathbf{Q}(\alpha+\pi, \beta) = \mathbf{Q}(\alpha, \beta + \pi/2)$. Second, the honeycomb AFM magnetic symmetry $\mathcal{M}_x\mathcal{T}$ imposes the mirror constraint $\mathbf{Q}(\alpha, \beta) = M_x \mathbf{Q}(\alpha, \pi/2 - \beta)$. When the illuminated system preserves $\mathcal{M}_x\mathcal{T}$ at $\beta = \pi/4$ and $3\pi/4$ (black arrows in Fig.~\ref{fig2}(b)), $\mathbf{Q}$ is strictly restricted along the $y$-axis ($Q_x = 0$). These two symmetry constraints provide a comprehensive all-optical scheme to manipulate $\mathbf{Q}$ or $\mathbf{P}$. Specifically:
\begin{itemize}
\item[(I)] \textit{All-optical reflection:} rotating the polarizer as $\beta \rightarrow \pi/2 - \beta$ performs a spatial reflection of the polar vector, reversing $Q_x \rightarrow -Q_x$;
\item[(II)] \textit{All-optical switching:} at the pinned orientation $\beta = \pi/4$, shifting the optical phase as $\alpha \rightarrow \alpha + \pi$ (flipping the light helicity) directly reverses the AFE polarization direction ($\mathbf{Q} \rightarrow -\mathbf{Q}$), as governed by the combined relation $\mathbf{Q}(\alpha+\pi, \beta) = M_x \mathbf{Q}(\alpha, -\beta)$.
\end{itemize}
This establishes a robust mechanism for all-optical, non-volatile writing of compensated polar states.

\begin{figure}[t]
    \centering
    \includegraphics[width=\linewidth]{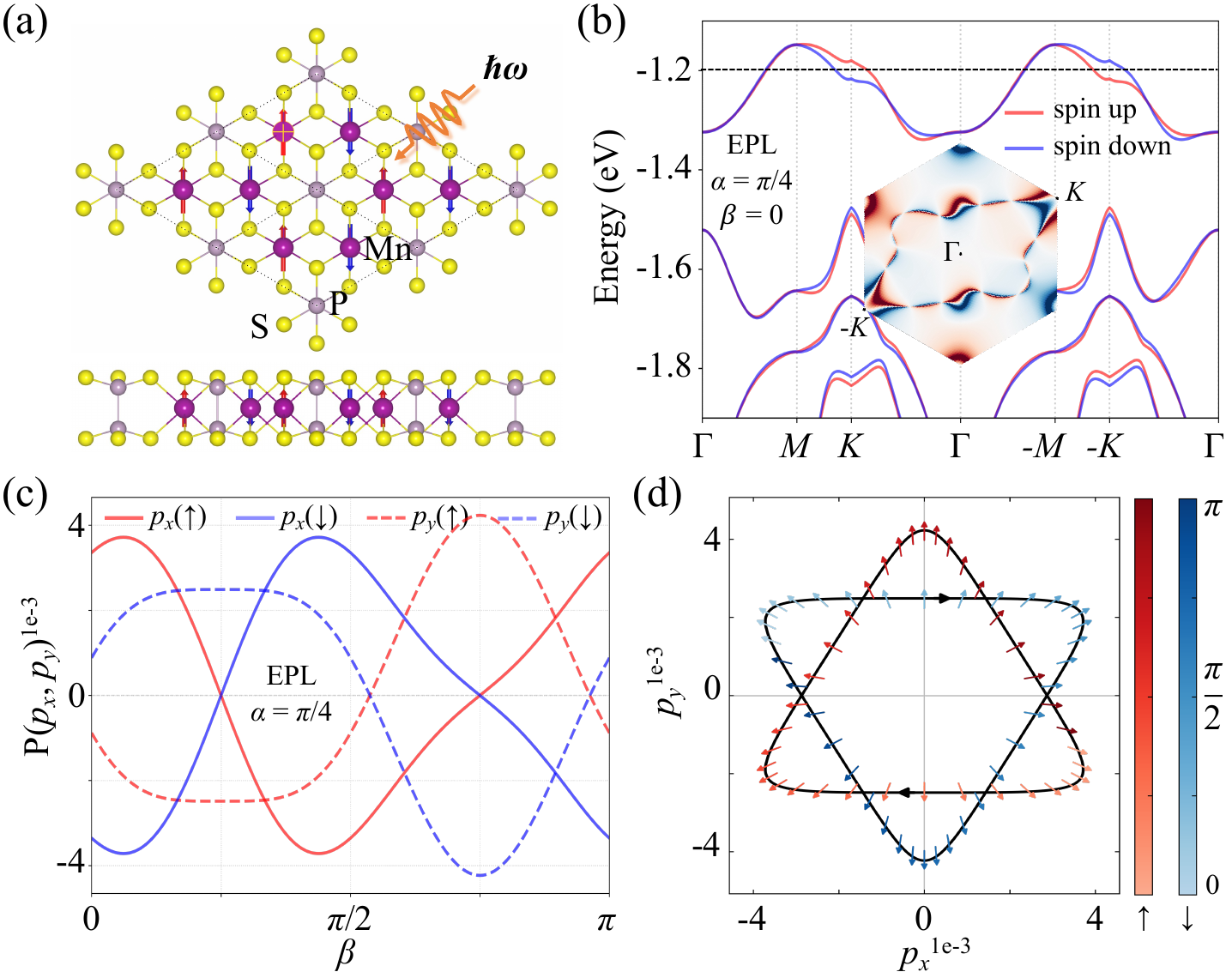}
    \caption{Floquet spin-antiferroelectricity in monolayer MnPS$_3$.
   (a) Crystal and magnetic structure of monolayer $\text{MnPS}_3$, where red and blue arrows represent the spin-up and spin-down magnetic moments on Mn sites. 
    (b) Spin-resolved band structure under EPL. The inset displays the constant-energy contours at $E=-1.2$~eV.
    (c) Spin-resolved electric polarizations $P_x$ and $P_y$ as functions of the EPL in-plane orientation angle $\beta$. Red and blue curves denote the two opposite spin sectors and display compensated values.
    (d) Parametric polarization trajectories in the $P_x$--$P_y$ plane. The black arrows indicate the direction of increasing $\beta$, whereas the red and blue arrows show the instantaneous polarization directions of the spin-up and spin-down sectors, respectively. The two spin-resolved loops are related by inversion through the origin, indicating finite sector polarizations with a vanishing total macroscopic polarization. In panels (b)--(d), the light amplitude $eA_0/\hbar=0.3~\mathrm{\mathring{A}}^{-1}$ and the photon energy $\hbar\omega=20$~eV are adopted.}
    \label{fig3}
\end{figure}

\textit{\textcolor{blue}{Material Realization.}}---To demonstrate the experimental feasibility of our proposal, we perform first-principles density functional theory (DFT) calculations on monolayer $\text{MnPS}_3$ (with computational details provided in the SM~\cite{SM}). Monolayer $\text{MnPS}_3$ belongs to the extensively studied transition-metal phosphorus trichalcogenide family and has been isolated and spectroscopically investigated down to the monolayer limit~\cite{Chittari2016MPX3,Long2017,Strasdas2023}. As shown in Fig.~\ref{fig3}(a), the $\text{Mn}^{2+}$ magnetic ions form a honeycomb lattice with collinear N\'eel antiferromagnetic order. Under an off-resonant EPL drive with $\alpha = \pi/4$, $eA_0/\hbar=0.3~\mathrm{\mathring{A}}^{-1}$ and $\hbar\omega=20$~eV, the time-reversal symmetry breaking induced by the rotating field generates a momentum-odd dynamical mass $W(\boldsymbol{k})$, which lifts the spin degeneracy. This yields an odd-parity spin splitting with a valley spin splitting of approximately $20~\mathrm{meV}$ at the $K$ and $-K$ valleys [Fig.~\ref{fig3}(b)]. Concurrently, the spatial rotation symmetry broken by EPL induces a compensated in-plane sector polarization, $P_x$ and $P_y$, reaching a peak magnitude of approximately $4.23\times10^{-3}$ in units of the in-plane polarization quantum [Fig.~\ref{fig3}(c)]. The strict compensation ($\mathbf{P}_\uparrow = -\mathbf{P}_\downarrow$) is mapped as parametric polarization trajectories in the $P_x-P_y$ plane [Fig.~\ref{fig3}(d)], where the spin-up and spin-down loops are related by inversion through the origin, guaranteeing a vanishing net macroscopic polarization ($\mathbf{P}_{\text{total}} = 0$). While monolayer $\text{MnPS}_3$ under EPL serves here as a representative example, its corresponding polarization responses under CPL and LPL, along with further candidate 2D AFMs exhibiting similar Floquet spin-antiferroelectricity, are detailed in the SM~\cite{SM}.

\textit{\textcolor{blue}{Conclusions and discussion}}---In summary, we have proposed a theoretical scheme for realizing non-equilibrium Floquet spin-antiferroelectricity in 2D collinear antiferromagnets. By utilizing symmetry arguments and a tight-binding model, we demonstrate that periodic optical driving successfully bypasses the conventional in-plane polarization constraints imposed by a vertical rotation axis ($\mathcal{C}^z_{n \ge 2}$). Specifically, while LPL only induces a compensated spin-antiferroelectric order, EPL enables the simultaneous coexistence of compensated in-plane sector polarizations and odd-parity spin splitting. This non-equilibrium multiferroic state is highly tunable via the laser polarization angle and phase, and its physical feasibility is validated in monolayer $\text{MnPS}_3$ through first-principles calculations.

The Floquet spin-antiferroelectric phase departs from currently known multiferroics in two fundamental aspects: (i) It enables the realization of in-plane electric polarizations perpendicular to the vertical rotation axis ($\mathcal{C}^z_{n \ge 2}$), a configuration conventionally forbidden by crystal symmetry; (ii) It allows this polar order to coexist with, and be synergistically controlled alongside, the unique unconventional magnetism. Notably, distinct from slow lattice-driven switching, the optically driven electronic mechanism enables ultrafast, non-destructive control of electric polarization~\cite{zhu2026light}. Furthermore, as detailed in the SM, the Floquet spin-antiferroelectric order can be effectively manipulated not only by altering the light polarization but also by engineering the lattice symmetry. In addition, we extend this Floquet engineering to altermagnetic systems, demonstrating that periodic driving can induce a spin-resolved macroscopic polarization with a non-zero net sum ($\mathbf{P}_\uparrow + \mathbf{P}_\downarrow \neq 0$). These insights expand the horizon of dynamic multiferroicity in quantum materials, offering fresh opportunities for realizing power-efficient, light-controlled information storage.

\emph{\textcolor{blue}{Acknowledgments.}}---
This work was supported by the National Key Research and Development Program of the Ministry of Science and Technology of China (Grant No. 2025YFA1411303), the National Natural Science Foundation of China (92365101, 12347101, 12074108, 12447141, 12404045, and 12474151), the Natural Science Foundation of Chongqing (2023NSCQ-JQX0024 and CSTB2022NSCQ-MSX0568).

\nocite{Fei2018WTe2,Meng2022Sliding,Hohenberg1964,Kohn1965,Bloechl1994,Kresse1996VASP,KresseJoubert1999,Perdew1996,Dudarev1998,Mostofi2008,Wu2018,Yang2023Cr2CCl2,Sattar2022MnX,Coak2019VPS3,Graf1995,Higashitarumizu2020SnS,Kruse2023TwoDimensional,Sambe1973,Shirley1965,Xu2020In2Se3,Akgenc2020Ti2C,Li2018Mg3X2,Limbu2025Cr2C,Ni2021MnPSe3,OkaKitamura2019,Pan2018Mg3C2,Vatansever2024Fe2O3}

\bibliography{ref}

\end{document}

\author{Yu-Hao Wei}
\affiliation{Institute for Structure and Function $\&$ Department of Physics $\&$ Chongqing Key Laboratory for Strongly Coupled Physics, Chongqing University, Chongqing 400044, People's Republic of China}
\author{Zheng Qin}
\affiliation{Institute for Structure and Function $\&$ Department of Physics $\&$ Chongqing Key Laboratory for Strongly Coupled Physics, Chongqing University, Chongqing 400044, People's Republic of China}
\author{Shengpu Huang}
\affiliation{Institute for Structure and Function $\&$ Department of Physics $\&$ Chongqing Key Laboratory for Strongly Coupled Physics, Chongqing University, Chongqing 400044, People's Republic of China}
\author{Dong-Hui Xu}
\affiliation{Institute for Structure and Function $\&$ Department of Physics $\&$ Chongqing Key Laboratory for Strongly Coupled Physics, Chongqing University, Chongqing 400044, People's Republic of China}
\affiliation{Center of Quantum materials and devices, Chongqing University, Chongqing 400044, People's Republic of China}
\author{Da-Shuai Ma}
\email{mads@cqu.edu.cn}
\affiliation{Institute for Structure and Function $\&$ Department of Physics $\&$ Chongqing Key Laboratory for Strongly Coupled Physics, Chongqing University, Chongqing 400044, People's Republic of China}
\affiliation{Center of Quantum materials and devices, Chongqing University, Chongqing 400044, People's Republic of China}
\author{Rui Wang}
\email{rcwang@cqu.edu.cn}
\affiliation{Institute for Structure and Function $\&$ Department of Physics $\&$ Chongqing Key Laboratory for Strongly Coupled Physics, Chongqing University, Chongqing 400044, People's Republic of China}
\affiliation{Center of Quantum materials and devices, Chongqing University, Chongqing 400044, People's Republic of China}
\affiliation{State Key Laboratory of Mechanical Transmission for Advanced Equipment, Chongqing University, Chongqing 400044, People's Republic of China}

\title{Supplemental Material for ``Floquet Spin-Antiferroelectricity in Collinear Antiferromagnets''
}
\maketitle

This Supplemental Material is organized as follows. Section I presents additional results for the monolayer, distorted, and bilayer honeycomb antiferromagnets and the four-site tetragonal altermagnet. Section II summarizes the first-principles, Wannier, and Floquet computational methods. Section III presents the Floquet band structures and spin-resolved polarization responses of monolayer MnPS$_3$ and MnPSe$_3$. Section IV lists additional two-dimensional antiferromagnetic material candidates. The final Supplementary Note provides the leading high-frequency analysis of the LPL-induced compensated polarization.

\section{Floquet Responses of Tight-Binding Models}

In this section, we provide additional results for the minimal honeycomb antiferromagnetic model discussed in the main text and examine the Floquet responses of three related tight-binding models. Specifically, we consider a uniaxially distorted honeycomb antiferromagnet, a bilayer honeycomb antiferromagnet, and a four-site tetragonal altermagnet. These models represent different lattice geometries and symmetry constraints and are used to compare how periodic optical driving modifies the spin splitting and spin-sector-resolved electric polarization. The model Hamiltonians, parameters, and corresponding band structures and polarization results are presented below.

\subsection{Monolayer Honeycomb Antiferromagnet}

\begin{figure}[H]
    \centering
    \includegraphics[width=0.63\linewidth]{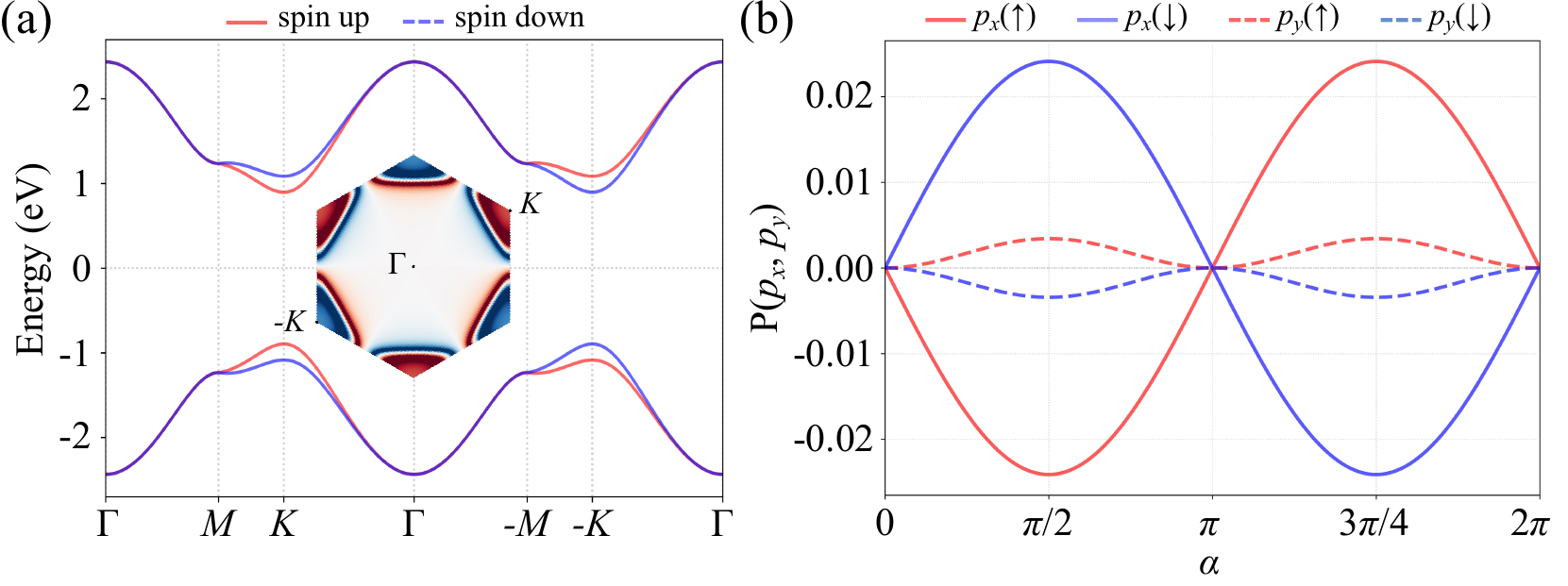}
    \caption{Floquet spin-antiferroelectricity in monolayer honeycomb antiferromagnet.
    (a) Spin-resolved Floquet band structures under CPL ($\alpha=0$), showing odd-parity spin splitting. Insets show spin-resolved isoenergy surfaces at $E = -1.0$~eV. (b) Spin-resolved electric polarization $P_x$ and $P_y$ as a function of the relative phase $\alpha$ at $\beta=0$. The polarization vanishes at the CPL points $\alpha=0$, $\pi$, and $2\pi$, whereas LPL and EPL allow finite compensated sector polarizations. The preserved threefold rotation forces $\mathbf P_\sigma=0$ at the CPL points, whereas LPL and EPL break this constraint and allow finite compensated polarizations satisfying $\mathbf P_\uparrow=-\mathbf P_\downarrow$.
    }
    \label{figS1}
\end{figure}

The equilibrium real-space Hamiltonian is written in the same notation as in the main text:
\begin{equation}
H_0 = -t\sum_{\langle ij\rangle}c_i^\dagger c_j+\mathrm{H.c.} +\sum_i m_i c_i^\dagger\sigma_z c_i,
\label{smeq12}
\end{equation}
where $c_i^\dagger=(c_{i\uparrow}^\dagger,c_{i\downarrow}^\dagger)$ is the electron creation spinor at site $i$ ($i\in A,B$), $t$ is the nearest-neighbor hopping amplitude, and $m_i$ is the local exchange field. The collinear N\'eel order corresponds to $m_A=-m_B\equiv m$. In the spin-major basis $(\phi_A^\uparrow,\phi_B^\uparrow,\phi_A^\downarrow,\phi_B^\downarrow)$, the Bloch Hamiltonian can be written as
\begin{equation}
H_0(\mathbf k)=\sigma_0\otimes
\begin{pmatrix}
0 & \Delta(\mathbf k) \\
\Delta^*(\mathbf k) & 0
\end{pmatrix}
+m\,\sigma_z\otimes\tau_z ,
\label{smeq13}
\end{equation}
where $\sigma_i$ and $\tau_i$ are Pauli matrices acting in the spin and sublattice spaces, respectively, and $\sigma_0$ is the identity matrix in spin space. Here,
$\Delta(\mathbf k)=-t\sum_{\ell=1}^{3}e^{i\mathbf k\cdot\boldsymbol{\delta}_\ell}$,
with $\boldsymbol{\delta}_{1,2,3}$ denoting the three nearest-neighbor bond vectors.

\subsection{Uniaxially Distorted Honeycomb Antiferromagnet}

\begin{figure}[H]
    \centering
    \includegraphics[width=0.75\linewidth]{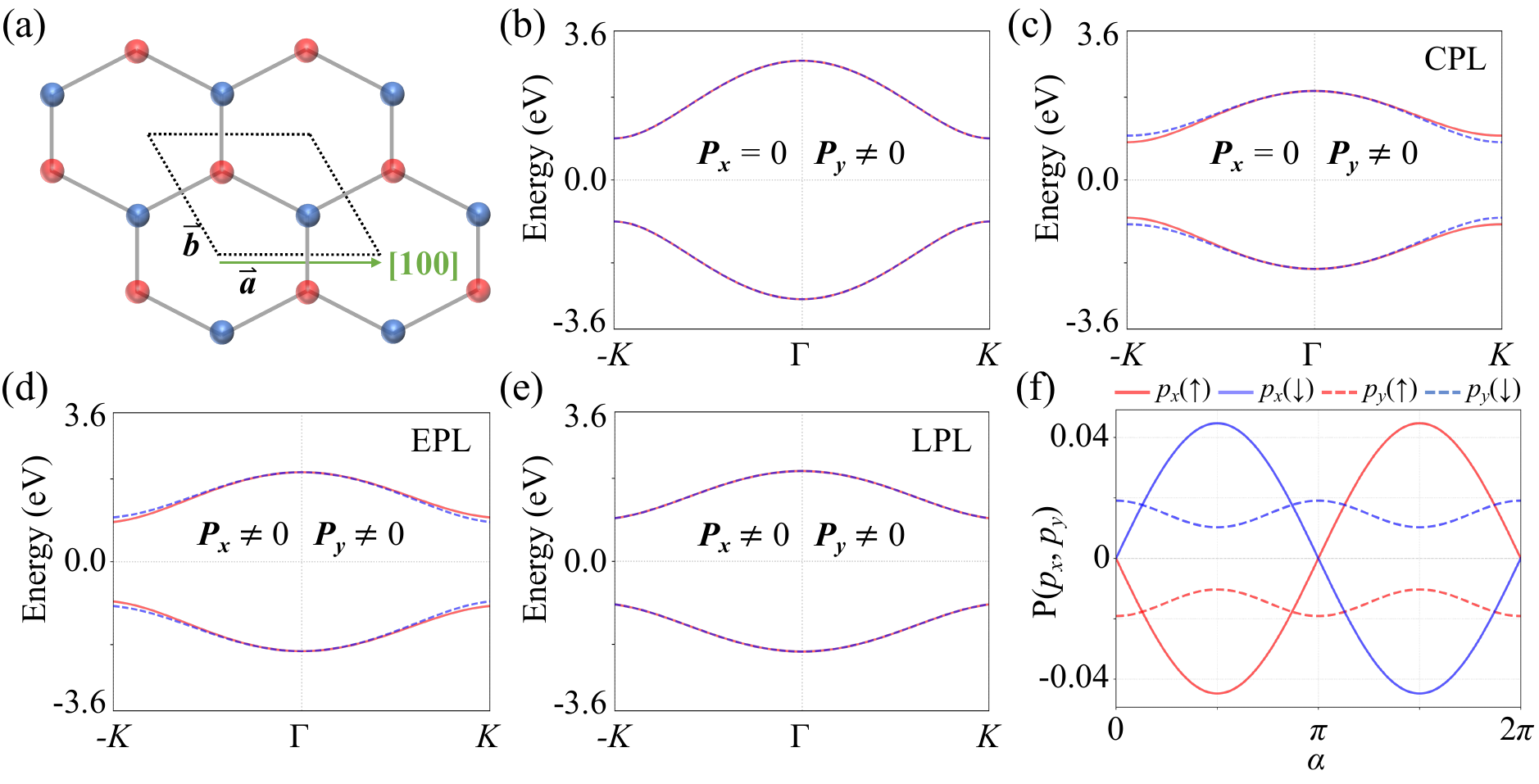}
    \caption{Floquet spin-antiferroelectricity in uniaxially distorted honeycomb antiferromagnet.
    (a) Lattice distorted by $15\%$ along the Cartesian $x$ direction, with the $[100]$ direction indicated. (b)--(e) Spin-resolved band structures of equilibrium and under CPL, EPL, and LPL, respectively. (f) Spin-resolved electric polarization components $P_x$ and $P_y$ as functions of the relative phase $\alpha$ at $\beta=0$. The residual mirror symmetry $\mathcal{M}_x$ enforces $P_x=0$ while allowing $P_y\neq0$, whereas CPL, EPL, and LPL modify the spin-resolved band structures and the $\alpha$-dependent sector polarization without restoring the threefold rotation broken by the uniaxial distortion.
    }
    \label{figS2}
\end{figure}

The distorted honeycomb model is obtained from the monolayer model above by applying a $15\%$ uniaxial tensile strain along the Cartesian $x$ direction. Each unit cell still contains one spinful $s$ orbital on each of the antiferromagnetic sublattices $A$ and $B$, and the staggered exchange term is unchanged. Only the bond vectors and their associated hopping amplitudes become direction dependent. The equilibrium real-space Hamiltonian is therefore
\begin{equation}
H_{\mathrm{dis}}
={}
\sum_{\mathbf R}\sum_{j=1}^{3}
t_j
c_{A\mathbf R}^{\dagger}
c_{B,\mathbf R+\mathbf R_j}
+
\mathrm{H.c.}
+
m\sum_{\mathbf R}
\left(
c_{A\mathbf R}^{\dagger}\sigma_z c_{A\mathbf R}
-
c_{B\mathbf R}^{\dagger}\sigma_z c_{B\mathbf R}
\right),
\label{smeq_distorted_real}
\end{equation}
where all symbols have the same meaning as in Eq.~\eqref{smeq12}, except that $t_j$ is now the hopping amplitude along the $j$th distorted bond. We take $m=1$~eV, $t_1=t_2=-0.85$~eV, and $t_3=-1$~eV. The equality $t_1=t_2$ follows from the remaining mirror symmetry, while $t_3\neq t_{1,2}$ describes the uniaxial distortion. In the spin-major basis $(\phi_A^\uparrow,\phi_B^\uparrow,\phi_A^\downarrow,\phi_B^\downarrow)$, the Bloch Hamiltonian can be written as
\begin{equation}
H_{\mathrm{dis}}(\mathbf k)
=
\sigma_0\otimes
\begin{pmatrix}
0 & \Delta_{\mathrm{dis}}(\mathbf k) \\
\Delta_{\mathrm{dis}}^*(\mathbf k) & 0
\end{pmatrix}
+
m\,\sigma_z\otimes\tau_z ,
\label{smeq_distorted_k}
\end{equation}
where $\sigma_i$ and $\tau_i$ are Pauli matrices acting in the spin and sublattice spaces, respectively, and $\sigma_0$ is the identity matrix in spin space. Here $\Delta_{\mathrm{dis}}(\mathbf k)=\sum_{j=1}^{3}t_j e^{i\mathbf k\cdot\boldsymbol{\delta}_j}$, with $\boldsymbol{\delta}_j=\mathbf R_j+\boldsymbol{\tau}_B-\boldsymbol{\tau}_A$ denoting the distorted nearest-neighbor bond vectors. The bond-dependent hopping amplitudes break the threefold rotational symmetry of the undistorted lattice, allowing a finite electric polarization within each spin sector already in equilibrium.

\subsection{Bilayer Honeycomb Antiferromagnet}

The bilayer honeycomb model contains four sites $(A_1,B_1,A_2,B_2)$ in each unit cell, where the subscripts label the two layers, and each site hosts one spinful $s$ orbital. Its equilibrium real-space Hamiltonian is given by
\begin{equation}
H_{\mathrm{BL}}=t\sum_{l=1}^{2}\sum_{\langle ij\rangle}c_{il}^{\dagger}c_{jl}+t_{\perp}\sum_{i=A,B}c_{i1}^{\dagger}c_{i2}+\mathrm{H.c.}+\sum_{i,l}m_{il}c_{il}^{\dagger}\sigma_zc_{il},
\label{smeq_bilayer_real}
\end{equation}
where $t=-1$~eV is the intralayer nearest-neighbor hopping amplitude, $t_{\perp}=-0.25$~eV is the vertical hopping between identical sublattices, and $m_{il}$ is the local exchange field. The two layers carry opposite N\'eel configurations, with $m_{A_1}=m_{B_2}=m$ and $m_{B_1}=m_{A_2}=-m$, where $m=1$~eV. In the spin-major basis $(\phi_{A_1}^{\uparrow},\phi_{B_1}^{\uparrow},\phi_{A_2}^{\uparrow},\phi_{B_2}^{\uparrow},\phi_{A_1}^{\downarrow},\phi_{B_1}^{\downarrow},\phi_{A_2}^{\downarrow},\phi_{B_2}^{\downarrow})$, the Bloch Hamiltonian can be written as
\begin{equation}
H_{\mathrm{BL}}(\mathbf k)=\sigma_0\otimes\left[\lambda_0\otimes\begin{pmatrix}0&\Delta(\mathbf k)\\ \Delta^*(\mathbf k)&0\end{pmatrix}+t_{\perp}\lambda_x\otimes\tau_0\right]+m\,\sigma_z\otimes\lambda_z\otimes\tau_z ,
\label{smeq_bilayer_k}
\end{equation}
where $\sigma_i$, $\tau_i$, and $\lambda_i$ are Pauli matrices acting in the spin, sublattice, and layer spaces, respectively, and $\sigma_0$, $\tau_0$, and $\lambda_0$ are the corresponding identity matrices. Here $\Delta(\mathbf k)=t\sum_{j=1}^{3}e^{i\mathbf k\cdot\boldsymbol{\delta}_j}$, with $\boldsymbol{\delta}_{1,2,3}$ denoting the three intralayer nearest-neighbor bond vectors.

\begin{figure}[H]
    \centering
    \includegraphics[width=0.8\linewidth]{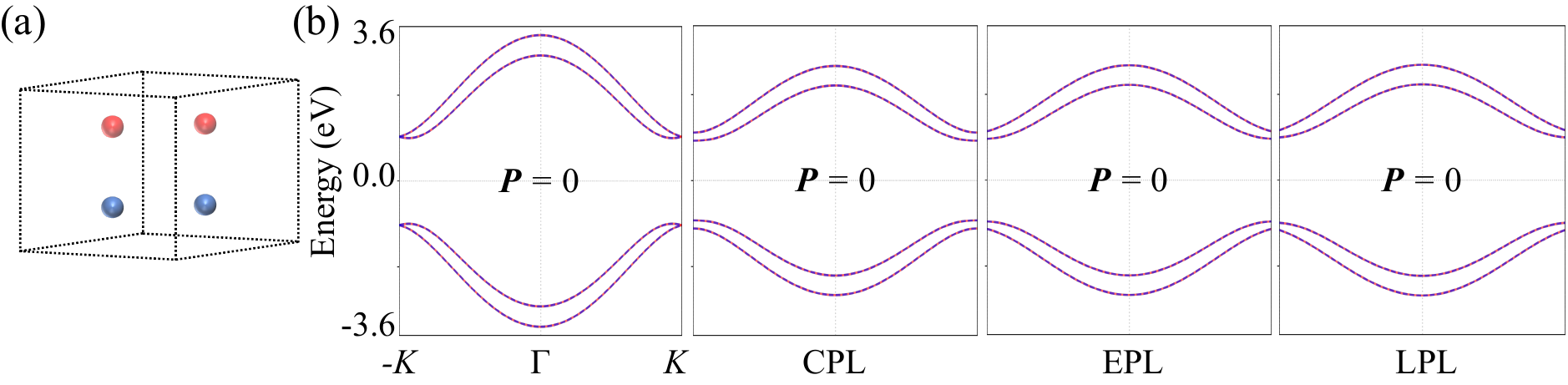}
    \caption{Floquet responses of the bilayer honeycomb antiferromagnet.
    (a) Bilayer lattice and magnetic configuration where the red and blue spheres denote opposite local exchange fields. (b) Spin-resolved band structures in equilibrium and under CPL, EPL, and LPL. The two spin sectors remain degenerate and the electric polarization remains zero in all four cases.
    }
    \label{figS3}
\end{figure}

As shown in Fig.~\ref{figS3}, the layer-exchange symmetry retains spin degeneracy and a vanishing electric polarization in equilibrium and under each of the three representative drives.

\subsection{Four-Site Tetragonal Altermagnet}

The tetragonal model is defined on a square lattice with primitive vectors $\mathbf a_1=a\hat{\mathbf x}$ and $\mathbf a_2=a\hat{\mathbf y}$. Each unit cell contains four spinful $s$ orbitals at the fractional coordinates $\boldsymbol{\tau}_{A_1}=(1/4,1/2)$, $\boldsymbol{\tau}_{A_2}=(3/4,1/2)$, $\boldsymbol{\tau}_{B_1}=(1/2,1/4)$, and $\boldsymbol{\tau}_{B_2}=(1/2,3/4)$. The $A$ and $B$ sites carry opposite local exchange fields and occur in equal numbers, giving a compensated collinear magnetic state. Defining the local spinor $c_{\mu\mathbf R}=(c_{\mu\mathbf R\uparrow},c_{\mu\mathbf R\downarrow})^{\mathsf T}$ and retaining the nearest- and next-nearest-neighbor shells, the equilibrium real-space Hamiltonian is
\begin{equation}
\begin{aligned}
H_{\mathrm{AM}}={}&
M\sum_{\mathbf R}\sum_{i=1}^{2}
\left(-c_{A_i\mathbf R}^{\dagger}\sigma_zc_{A_i\mathbf R}
+c_{B_i\mathbf R}^{\dagger}\sigma_zc_{B_i\mathbf R}\right)\\
&+\sum_{\mathbf R}\Bigg[
t_1\sum_{i,j=1}^{2}c_{A_i\mathbf R}^{\dagger}\sigma_0c_{B_j\mathbf R}\\
&\quad+c_{A_1\mathbf R}^{\dagger}T_A^{(0)}c_{A_2\mathbf R}
+c_{A_1\mathbf R}^{\dagger}T_A^{(x)}c_{A_2,\mathbf R-\mathbf a_1}\\
&\quad+c_{B_1\mathbf R}^{\dagger}T_B^{(0)}c_{B_2\mathbf R}
+c_{B_1\mathbf R}^{\dagger}T_B^{(y)}c_{B_2,\mathbf R-\mathbf a_2}
+\mathrm{H.c.}\Bigg],
\end{aligned}
\label{smeq_alter_real}
\end{equation}
where $\sigma_i$ act in spin space and
\begin{equation}
\begin{aligned}
T_A^{(0)}&=\bar t_0\sigma_0+\delta t_0\sigma_z,&
T_B^{(0)}&=\bar t_0\sigma_0-\delta t_0\sigma_z,\\
T_A^{(x)}&=\bar t_2\sigma_0+\delta t_2\sigma_z,&
T_B^{(y)}&=\bar t_2\sigma_0-\delta t_2\sigma_z,
\end{aligned}
\label{smeq_alter_hoppings}
\end{equation}
with $\bar t_0=(t_A^{(0)}+t_B^{(0)})/2$, $\delta t_0=(t_A^{(0)}-t_B^{(0)})/2$, $\bar t_2=(t_A^{(x)}+t_B^{(y)})/2$, and $\delta t_2=(t_A^{(x)}-t_B^{(y)})/2$. Thus, the spin-dependent directional hopping is written explicitly in Eq.~\eqref{smeq_alter_real}. Under $\mathcal S=[C_{2,\mathrm{spin}}\parallel C_{4z}]$, the two spin projections and the $A/B$ hopping patterns are interchanged, with $(A_1,A_2,B_1,B_2)\mapsto(B_1,B_2,A_2,A_1)$. In the spin-major basis $\Psi=(\phi_{A_1}^\uparrow,\phi_{A_2}^\uparrow,\phi_{B_1}^\uparrow,\phi_{B_2}^\uparrow,\phi_{A_1}^\downarrow,\phi_{A_2}^\downarrow,\phi_{B_1}^\downarrow,\phi_{B_2}^\downarrow)^{\mathsf T}$, the Bloch Hamiltonian can be written as
\begin{equation}
H_{\mathrm{AM}}(\mathbf k)
=
\begin{pmatrix}
h_\uparrow(\mathbf k)&0\\
0&h_\downarrow(\mathbf k)
\end{pmatrix},
\qquad
h_\downarrow(\mathbf k)
=U_{C_{4z}}h_\uparrow(C_{4z}^{-1}\mathbf k)U_{C_{4z}}^\dagger,
\label{smeq_alter_k}
\end{equation}
where the spin-up block is
\begin{equation}
h_\uparrow(\mathbf k)=
\begin{pmatrix}
-M&f_A(k_x)&t_1&t_1\\
f_A^*(k_x)&-M&t_1&t_1\\
t_1&t_1&M&f_B(k_y)\\
t_1&t_1&f_B^*(k_y)&M
\end{pmatrix}.
\label{smeq_alter_hup}
\end{equation}
Here $f_A(k_x)=t_A^{(0)}+t_A^{(x)}e^{-ik_xa}$ and $f_B(k_y)=t_B^{(0)}+t_B^{(y)}e^{-ik_ya}$. We use $M=0.50$~eV, $t_1=0.82$~eV, $t_A^{(0)}=0.56$~eV, $t_A^{(x)}=-0.33$~eV, $t_B^{(0)}=-0.35$~eV, and $t_B^{(y)}=-0.32$~eV. The opposite spin sectors are therefore related by $E_\uparrow(\mathbf k)=E_\downarrow(C_{4z}\mathbf k)$ but are generally nondegenerate at the same generic momentum. Together with the compensated magnetic configuration, this spin-sector-switching relation identifies the model as a tetragonal altermagnet.

\begin{figure}[H]
    \centering
    \includegraphics[width=0.75\linewidth]{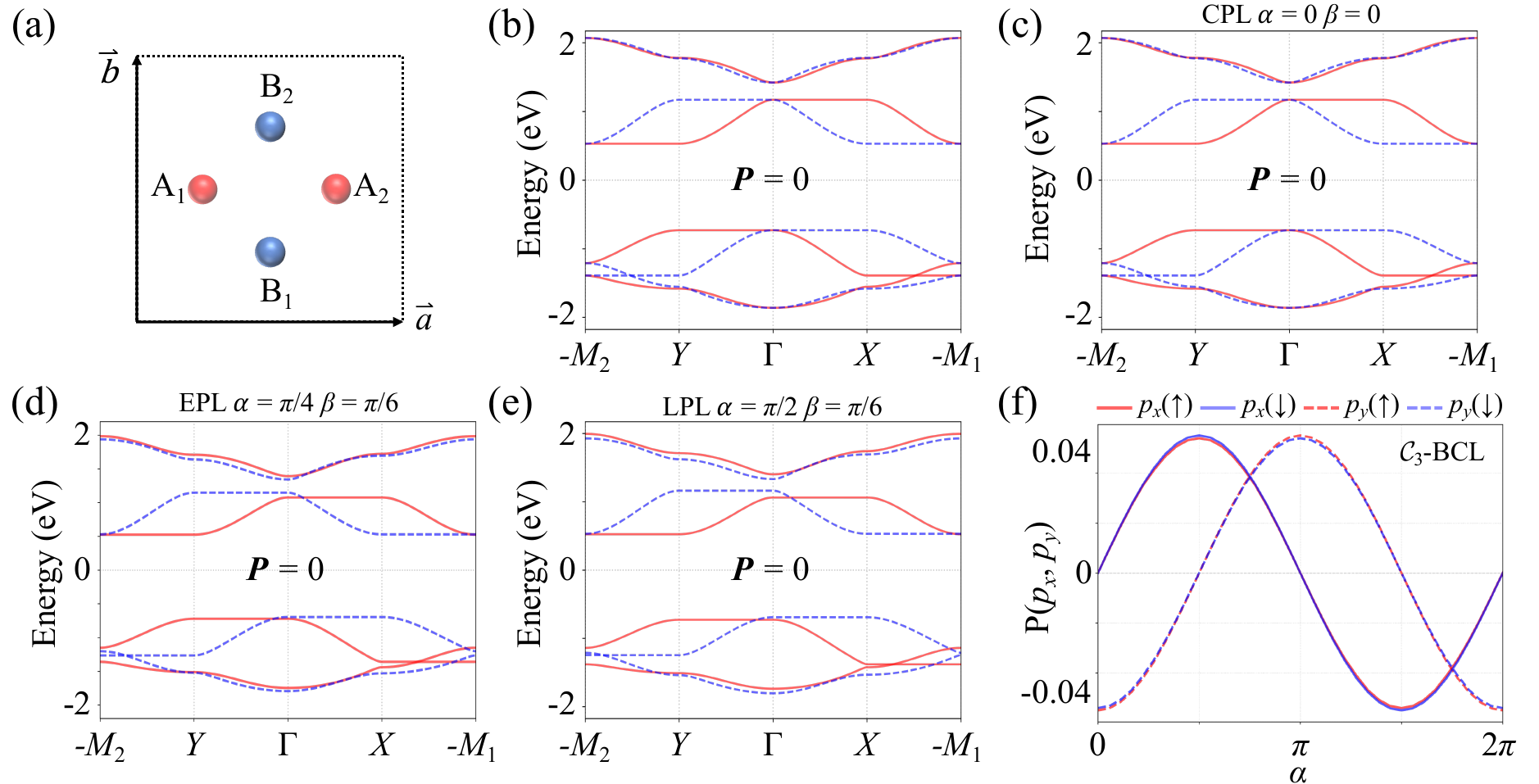}
    \caption{Floquet responses of the four-site tetragonal altermagnet.
    (a) Four-site magnetic unit cell where the red and blue spheres denote the two opposite magnetic sublattices. (b)--(e) Spin-resolved band structures of equilibrium and under CPL, EPL, and LPL, respectively. (f) Spin-resolved electric polarization $P_x$ and $P_y$ induced by threefold-symmetric trefoil bicircularly polarized light ($\mathcal{C}_3$-BCL) as functions of the relative phase $\alpha$ between its two counter-rotating circularly polarized components. Under the $\mathcal{C}_3$-BCL drive, the two spin sectors acquire closely spaced but distinguishable in-plane polarization curves, demonstrating a Floquet spin-ferroelectric state with a net electric polarization.
    }
    \label{figS4}
\end{figure}

\section{Calculation details}

First-principles calculations were performed within density functional theory (DFT)~\cite{Hohenberg1964,Kohn1965} using the Vienna \textit{Ab Initio} Simulation Package (VASP)~\cite{Kresse1996VASP}. The electron--ion interactions were described using the projector augmented-wave (PAW) method~\cite{Bloechl1994,KresseJoubert1999}, and the exchange--correlation functional was treated within the Perdew--Burke--Ernzerhof generalized gradient approximation (GGA-PBE)~\cite{Perdew1996}. The localized Mn $3d$ states were treated using the rotationally invariant GGA+$U$ method~\cite{Dudarev1998}, with an effective on-site Coulomb parameter $U_{\mathrm{eff}}=U-J=4$~eV. The plane-wave kinetic-energy cutoff was set to $500$~eV, and the Brillouin zone was sampled using an $18\times18\times1$ $\Gamma$-centered $\mathbf{k}$-point mesh. The self-consistent electronic iterations were converged to an energy tolerance of $10^{-7}$~eV. All atomic positions were relaxed until the residual Hellmann--Feynman forces were smaller than $10^{-2}$~eV/\AA.

To investigate the response under periodic optical driving, we constructed a tight-binding Hamiltonian based on maximally localized Wannier functions. The converged DFT Bloch states were projected onto localized Wannier orbitals using the Wannier90 package\cite{Mostofi2008,Wu2018}. In the absence of the optical field, the real-space tight-binding Hamiltonian is written as
\begin{equation}
H_0
=
\sum_{mn}\sum_{\mathbf R_j}
t_{mn}(\mathbf R_j)
c_m^\dagger(\mathbf R_j)
c_n(\mathbf R_0)
+
\mathrm{H.c.},
\label{smeq1}
\end{equation}
where $t_{mn}(\mathbf R_j)$ is the hopping amplitude from the $n$th Wannier orbital in the home unit cell to the $m$th Wannier orbital in the unit cell specified by the lattice vector $\mathbf R_j$. The operators $c_m^\dagger(\mathbf R_j)$ and $c_n(\mathbf R_0)$ are the corresponding fermionic creation and annihilation operators.

The optical field is introduced through the Peierls substitution \cite{Graf1995}, with the electron charge written as $-e$ ($e>0$). This convention corresponds to $\mathbf k\rightarrow\mathbf k-e\boldsymbol{\mathcal A}(t)/\hbar$ and gives
\begin{equation}
t_{mn}(\mathbf R_j)
\rightarrow
t_{mn}(\mathbf R_j,t)
=
t_{mn}(\mathbf R_j)
\exp\left[
-i \frac{e}{\hbar}\boldsymbol{\mathcal A}(t)\cdot
\mathbf d_{mn}(\mathbf R_j)
\right],
\label{smeq2}
\end{equation}
where $\boldsymbol{\mathcal A}(t)$ is the vector potential. The displacement vector between the two Wannier centers is
\begin{equation}
\mathbf d_{mn}(\mathbf R_j)
=
\mathbf R_j+\boldsymbol{\tau}_m-\boldsymbol{\tau}_n,
\label{smeq3}
\end{equation}
where $\boldsymbol{\tau}_m$ and $\boldsymbol{\tau}_n$ denote the intracell positions of the corresponding Wannier centers.

The time-periodic vector potential is parameterized as
\begin{equation}
\boldsymbol{\mathcal A}(t)
=
A_0
\left[
\mathbf e_1\cos(\omega t)
+
\mathbf e_2\sin(\omega t+\alpha)
\right],
\label{smeq4}
\end{equation}
where $A_0$ is the amplitude of the vector potential, $\omega$ is the driving frequency, and $\alpha$ controls the relative phase between the two components. The polarization basis vectors are defined as
\begin{equation}
\mathbf e_1
=
(\cos\beta,\sin\beta,0),
\qquad
\mathbf e_2
=
(-\sin\beta\cos\theta,\cos\beta\cos\theta,\sin\theta).
\label{smeq5}
\end{equation}
Here, $\beta$ determines the in-plane orientation of the polarization basis, while $\theta$ controls its out-of-plane tilting. For $\theta=0$ or $\pi$, both basis vectors lie in the $xy$ plane.

Within the convention of Eq.~\eqref{smeq4}, the phase difference between the two orthogonal field components is $\alpha-\pi/2$. Accordingly, $\alpha=0$, $\pi$, and $2\pi$ correspond to circularly polarized light, $\alpha=\pi/2$ and $3\pi/2$ correspond to linearly polarized light, and other values of $\alpha$ correspond to elliptically polarized light.

After the Peierls substitution, the time-dependent Bloch Hamiltonian takes the form
\begin{equation}
H(\mathbf k,t)
=
\sum_{mn}
H_{mn}(\mathbf k,t)
c_m^\dagger(\mathbf k)c_n(\mathbf k),
\label{smeq6}
\end{equation}
where
\begin{equation}
H_{mn}(\mathbf k,t)
=
\sum_{\mathbf R_j}
t_{mn}(\mathbf R_j)
e^{i\mathbf k\cdot\mathbf R_j}
\exp\left[
-i \frac{e}{\hbar}\boldsymbol{\mathcal A}(t)\cdot
\mathbf d_{mn}(\mathbf R_j)
\right].
\label{smeq7}
\end{equation}
Because the Hamiltonian is periodic in time,
$H(\mathbf k,t+T)=H(\mathbf k,t)$,
with $T=2\pi/\omega$, it can be represented as a time-independent matrix in the extended Floquet space \cite{Shirley1965,Sambe1973}. We define the Fourier coefficient of the photon-assisted hopping factor as
\begin{equation}
g_l\left[\mathbf d_{mn}(\mathbf R_j)\right]
=
\frac{1}{T}
\int_0^T
\exp\left[
-i \frac{e}{\hbar}\boldsymbol{\mathcal A}(t)\cdot
\mathbf d_{mn}(\mathbf R_j)
\right]
e^{-il\omega t}
dt,
\label{smeq8}
\end{equation}
where $l$ denotes the difference between two Floquet harmonic indices. The integral is evaluated numerically over one driving period.

The Floquet Hamiltonian in the extended photon space is then given by
\begin{equation}
\left[H_F(\mathbf k)\right]_{mn}^{pq}
=
\sum_{\mathbf R_j}
t_{mn}(\mathbf R_j)
g_{q-p}\left[\mathbf d_{mn}(\mathbf R_j)\right]
e^{i\mathbf k\cdot\mathbf R_j}
-
p\hbar\omega\,
\delta_{pq}\delta_{mn},
\label{smeq9}
\end{equation}
where $p$ and $q$ are Floquet harmonic indices, and the second term accounts for the photon-sector energy shift.

For the analytical discussion, we also employ the high-frequency expansion in the off-resonant regime~\cite{Goldman2014,Bukov2015,Eckardt2017,OkaKitamura2019},
\begin{equation}
H_{\mathrm{eff}}(\mathbf k)
=H_0(\mathbf k)
+\sum_{n\geq1}\frac{[H_{-n}(\mathbf k),H_n(\mathbf k)]}{n\hbar\omega}
+\mathcal O\!\left(\frac{1}{(\hbar\omega)^2}\right),
\label{smeq10}
\end{equation}
where $H_n(\mathbf k)=T^{-1}\int_0^T e^{-in\omega t}H(\mathbf k,t)\,dt$ and $H_0(\mathbf k)$ is the time-averaged Hamiltonian. Equation~\eqref{smeq10} is used only for the analytical derivation of the light-induced effective Hamiltonian and spin-resolved polarization. All numerical Floquet band structures and polarizations reported in this work are instead obtained by truncating Eq.~\eqref{smeq9} to $p,q=-N_F,\ldots,N_F$ and directly diagonalizing the resulting Sambe-space Hamiltonian.

\section{Floquet Responses of \texorpdfstring{$\mathrm{MnPX_3}$}{MnPX3} Monolayers}

At a broad phenomenological level, two-dimensional ferroelectrics can be organized into two major families. Intrinsic monolayer ferroelectrics develop a polar state within a single atomic layer, most commonly through an inversion-breaking lattice distortion accompanied by electronic charge redistribution. Stacking- or sliding-induced ferroelectrics instead require two or more layers and reverse their polarization by changing the interlayer registry. Although this classification is not exhaustive, it captures the dominant material platforms: interlayer sliding generally reverses a registry-dependent charge transfer and therefore predominantly produces an out-of-plane polarization~\cite{Fei2018WTe2,Meng2022Sliding}, whereas switchable ferroelectricity in the strict monolayer limit is dominated by in-plane polarization. Indeed, a high-throughput survey of dynamically stable insulating monolayers identified 49 ferroelectrics with purely in-plane polarization, compared with only 8 with purely out-of-plane polarization and 6 with mixed components~\cite{Kruse2023TwoDimensional}. An in-plane polar axis is particularly favorable at the two-dimensional limit because it produces no bound charge on the top and bottom surfaces, $\sigma_b=\mathbf P\cdot\hat{\mathbf z}=0$, thereby strongly suppressing the depolarization field and associated critical-thickness limitation; it also provides a vector degree of freedom that can be coupled directly to crystallographic anisotropy and lateral transport~\cite{Higashitarumizu2020SnS}. Nevertheless, intrinsic in-plane antiferroelectricity is substantially rarer: its first experimental realization in a two-dimensional material was reported only in nanostripe-ordered $\beta'$-$\mathrm{In_2Se_3}$, where neighboring stripes carry antiparallel local in-plane polarizations~\cite{Xu2020In2Se3}. The constraint is still sharper for type-II spin antiferroelectricity. Among the 18 compatible collinear spin point groups listed in Table~I of Ref.~\cite{Wang2026TypeII}, only four entries permit an in-plane component of $\mathbf Q$, with the allowed forms $(Q_1,Q_2,Q_3)$, $Q_2$, $(Q_1,0,Q_3)$, and $Q_1$; the remaining 14 constrain the order to $Q_3$. These material and symmetry restrictions motivate the present Floquet route: polarized light dynamically removes the in-plane nonpolar constraint without requiring either a rare equilibrium antipolar distortion or a multilayer sliding structure, thereby generating and continuously steering compensated in-plane electronic polarizations in broad classes of otherwise nonpolar, high-symmetry antiferromagnetic monolayers.

\begin{figure}[H]
    \centering
    \includegraphics[width=0.6\linewidth]{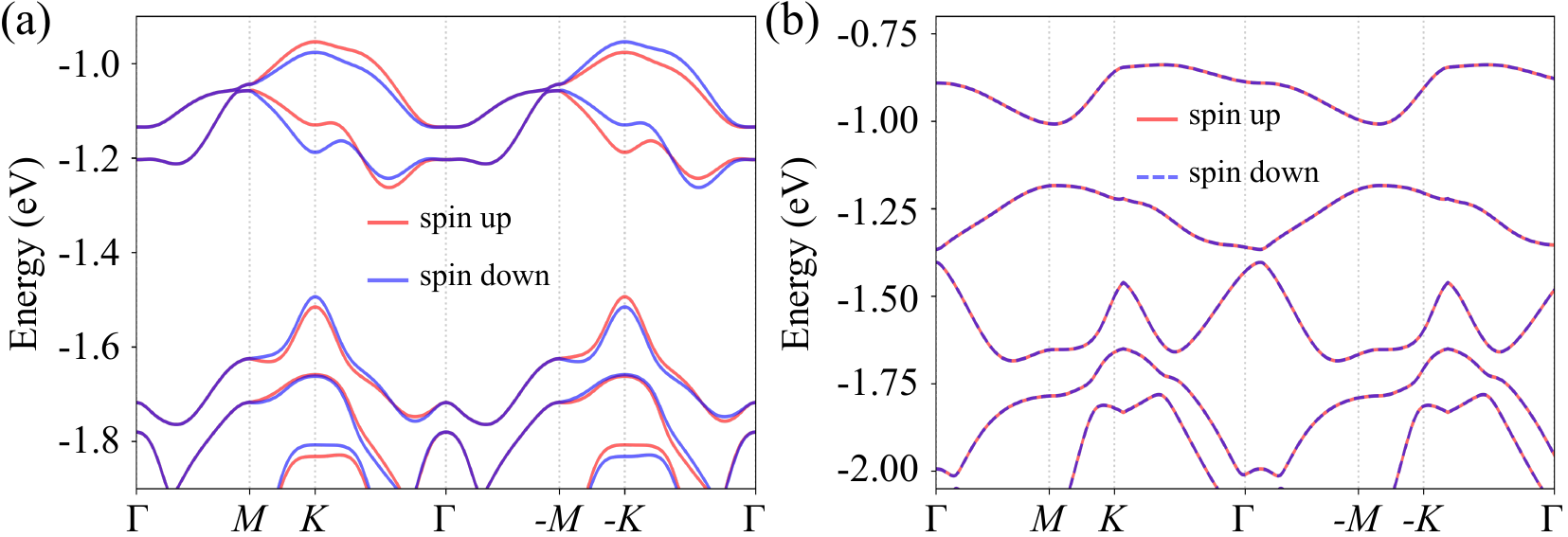}
    \caption{Spin-resolved Floquet band structures of monolayer $\mathrm{MnPS_3}$.
    (a) CPL ($\alpha=0$) produces odd-parity spin splitting. (b) LPL ($\alpha=\pi/2$) preserves spin degeneracy. Red solid and blue dashed curves denote the spin-up and spin-down sectors, respectively.
    }
    \label{figS5}
\end{figure}

\begin{figure}[H]
    \centering
    \includegraphics[width=0.57\linewidth]{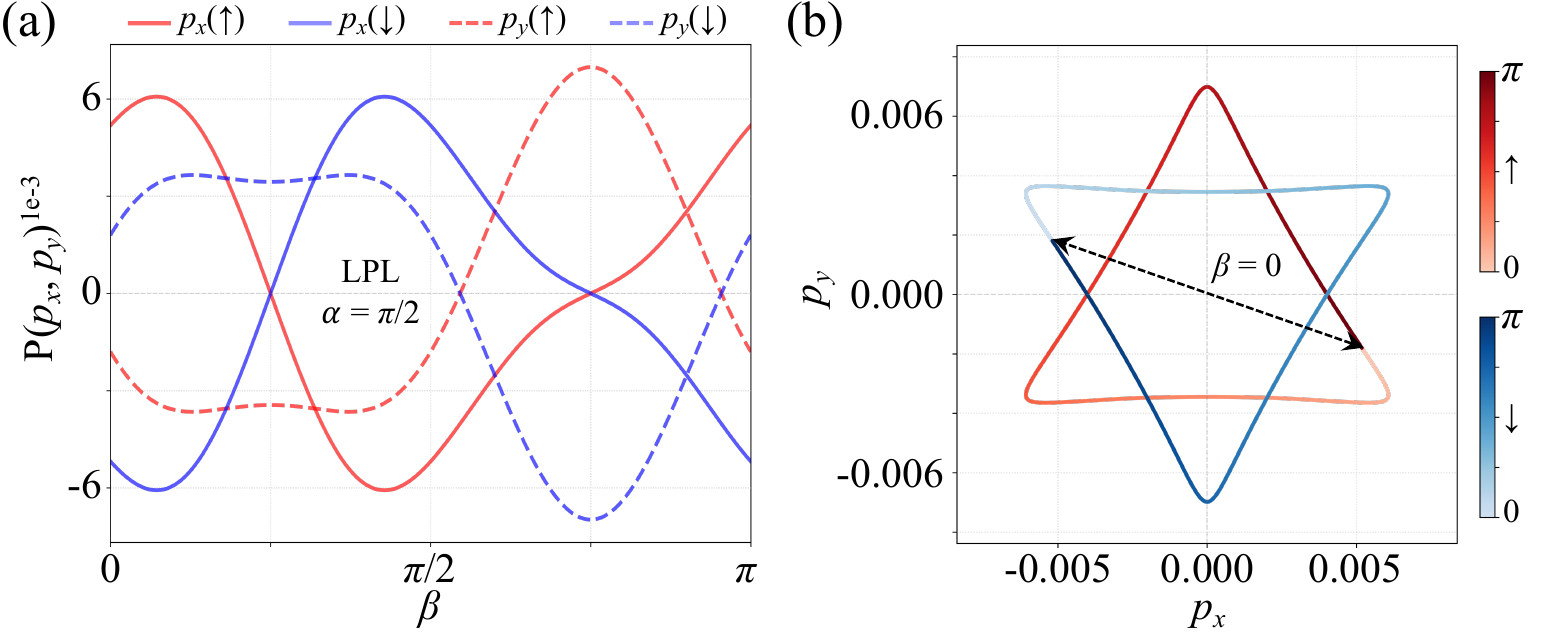}
    \caption{LPL-induced compensated polarization in monolayer $\mathrm{MnPS_3}$ for $\alpha=\pi/2$.
    (a) Spin-resolved polarization $P_x$ and $P_y$ as functions of the in-plane orientation angle $\beta$. (b) The corresponding parametric trajectories in the $(P_x,P_y)$ plane; the two spin-sector loops are related by inversion through the origin, demonstrating $\mathbf P_\uparrow=-\mathbf P_\downarrow$.
    }
    \label{figS6}
\end{figure}

\begin{figure}[H]
    \centering
    \includegraphics[width=0.85\linewidth]{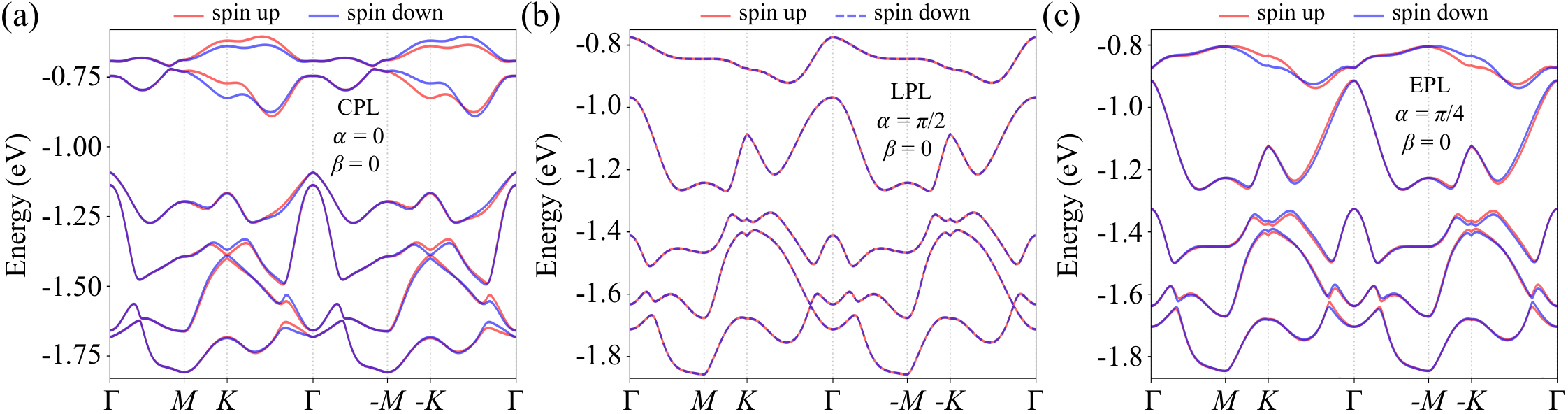}
    \caption{Spin-resolved Floquet band structures of monolayer $\mathrm{MnPSe_3}$ under (a) CPL with $\alpha=0$, (b) LPL with $\alpha=\pi/2$, and (c) EPL with $\alpha=\pi/4$. CPL and EPL produce spin splitting, whereas the LPL bands remain spin-degenerate. Red and blue curves denote the spin-up and spin-down sectors, respectively.
    }
    \label{figS7}
\end{figure}

\begin{figure}[H]
    \centering
    \includegraphics[width=0.57\linewidth]{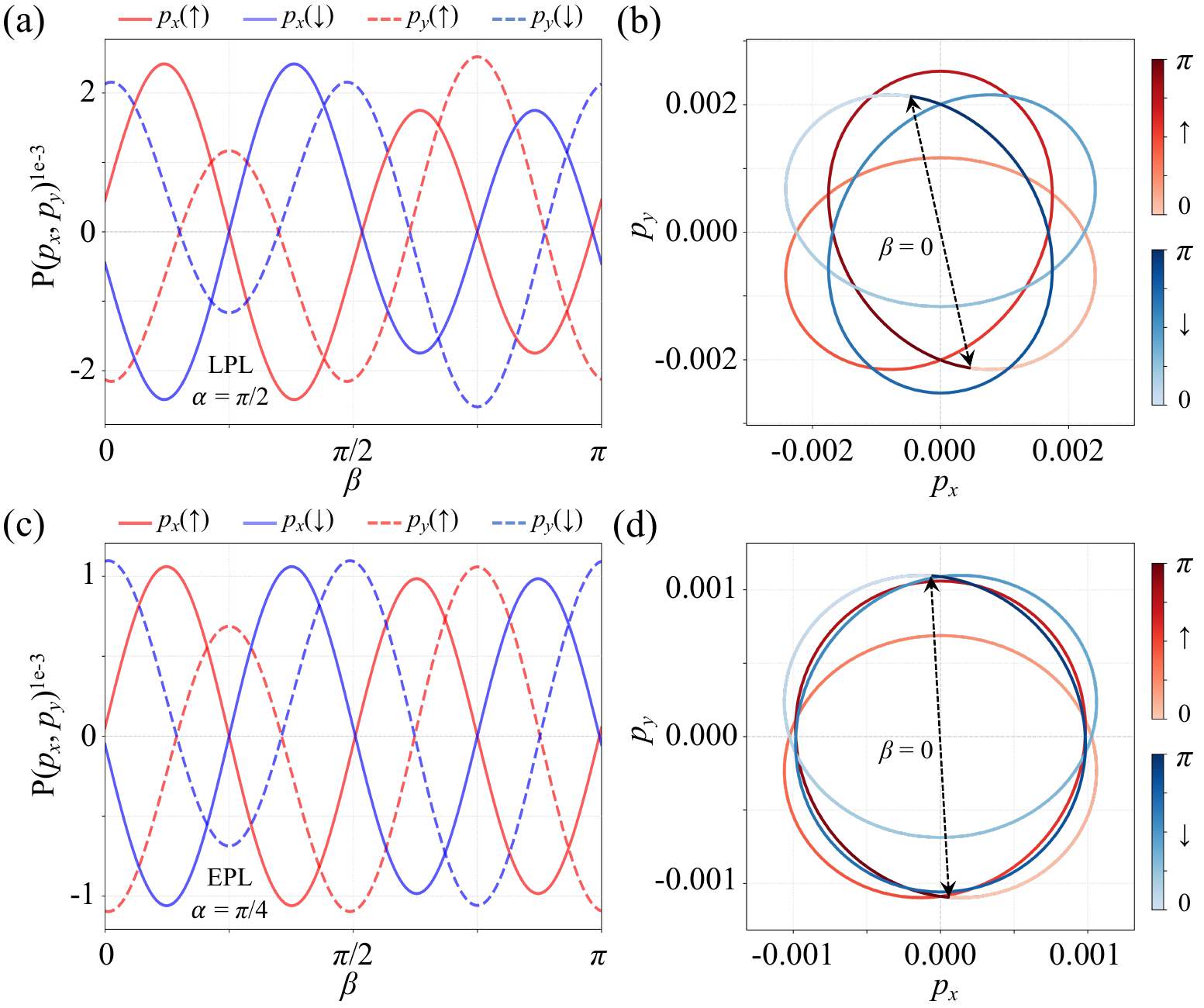}
    \caption{
    Spin-resolved electric polarization of monolayer $\mathrm{MnPSe_3}$. (a) $P_x$ and $P_y$ as functions of $\beta$ under LPL with $\alpha=\pi/2$. (b) Corresponding LPL trajectories in the $(P_x,P_y)$ plane. (c) $P_x$ and $P_y$ as functions of $\beta$ under EPL with $\alpha=\pi/4$. (d) Corresponding EPL trajectories in the $(P_x,P_y)$ plane. In both cases, the two spin sectors satisfy $\mathbf P_\uparrow=-\mathbf P_\downarrow$.
    }
    \label{figS8}
\end{figure}

Using the Wannier--Floquet framework described in Sec.~II, we calculate the spin-resolved band structures and polarizations of monolayer $\mathrm{MnPX_3}$ ($X=\mathrm{S},\mathrm{Se}$) under CPL, LPL, and EPL. For MnPS$_3$, CPL produces odd-parity spin splitting while the residual threefold rotation forbids in-plane sector polarization [Fig.~\ref{figS5}], whereas LPL preserves spin degeneracy but breaks the rotational constraint and generates finite compensated polarization with $\mathbf P_\uparrow=-\mathbf P_\downarrow$ [Fig.~\ref{figS6}]. MnPSe$_3$ exhibits the same symmetry-controlled behavior: CPL and EPL permit spin splitting, while LPL remains spin-degenerate [Fig.~\ref{figS7}], and both LPL and EPL induce finite compensated sector polarizations [Fig.~\ref{figS8}]. Thus, EPL realizes the coexistence of odd-parity spin splitting and compensated spin-antiferroelectric order in both material systems, consistent with the minimal-model analysis and the MnPS$_3$ results in the main text.

\section{Additional Two-Dimensional Antiferromagnetic Material Candidates}

Following the selection rules in the main text, candidate materials should host a collinear compensated AFM insulating state within a single structural layer and well-defined spin sectors in the nonrelativistic limit. Representative examples are the N\'eel-ordered $\mathrm{MPX_3}$ monolayers with the ideal trigonal layer structure $P\bar{3}1m$ (No.~162), including $\mathrm{MnPS_3}$ ($T_N\simeq78$~K in bulk), $\mathrm{MnPSe_3}$ ($T_N\simeq40$~K in the monolayer and approximately $74$~K in bulk), and the predicted semiconducting $\mathrm{VPS_3}$ and $\mathrm{VPSe_3}$ monolayers~\cite{Chittari2016MPX3,Huang2026LightOdd,Ni2021MnPSe3}. Ising-limit calculations at $U=4$~eV estimate ordering temperatures of approximately $570$ and $400$~K for $\mathrm{VPS_3}$ and $\mathrm{VPSe_3}$, respectively; these theoretical estimates are distinct from the approximately $62$~K transition measured in bulk vanadium-deficient $\mathrm{V_{0.9}PS_3}$~\cite{Chittari2016MPX3,Coak2019VPS3}. The tellurides $\mathrm{VPTe_3}$ and $\mathrm{MnPTe_3}$ provide additional predicted members of this family, although their transition temperatures have not yet been established~\cite{Chittari2016MPX3}. In all of these N\'eel structures, inversion exchanges the opposite-spin sublattices whereas the threefold rotation preserves each sublattice, reproducing the symmetry ingredients of the honeycomb model. A second family consists of the isostructural, symmetrically terminated $P\bar{3}m1$ (No.~164) MXenes $\mathrm{Cr_2CCl_2}$, $\mathrm{Cr_2CF_2}$, and $\mathrm{Cr_2C(OH)_2}$. They are A-type AFM semiconductors with inversion-related, oppositely magnetized Cr sublayers; consequently, they possess the same spin-group structure relevant to the light-induced compensated in-plane polarization. The predicted ordering temperature of $\mathrm{Cr_2CCl_2}$ is approximately $1300$~K, while transition temperatures have not been reported for the F- and OH-terminated compounds~\cite{Yang2023Cr2CCl2,Limbu2025Cr2C}. The same structural class also includes the predicted AFM semiconducting $1T$-$\mathrm{Ti_2C}$ monolayer, whose dynamical and thermal stability has been confirmed by phonon and ab initio molecular-dynamics calculations~\cite{Akgenc2020Ti2C}. Further candidates are $\mathrm{MnS}$, $\mathrm{MnSe}$, and $\mathrm{MnTe}$ in the $P\bar{3}m1$ structure, which combine an AFM semiconducting state with inversion-related magnetic sublayers~\cite{Sattar2022MnX}. Beyond transition-metal compounds, the predicted planar honeycomb--kagome monolayers $\mathrm{Mg_3C_2}$ and $\mathrm{Mg_3Si_2}$ are dynamically and mechanically stable AFM semiconductors, providing a distinct $p$-orbital realization of the same compensated two-sublattice setting~\cite{Li2018Mg3X2,Pan2018Mg3C2}. Another promising candidate is planar honeycomb--kagome $\mathrm{Fe_2O_3}$ (predicted $T_N=631.7(5)$~K), whose opposite-spin Fe sites are exchanged by inversion and individually preserved by threefold rotation~\cite{Vatansever2024Fe2O3}. Together, these chemically and structurally diverse families provide a broad materials platform for exploring and optimizing Floquet-induced compensated in-plane polarization.

\clearpage

\section*{Supplementary Note: Derivation of the Effective Floquet Hamiltonian and Spin-Resolved Polarizations under Linearly Polarized Light}

\subsection{General Description of the Periodic Driving Light Field}
To describe the interaction between the collinear antiferromagnetic (AFM) system and periodic driving, we introduce a general time-periodic vector potential:
\begin{equation}
\boldsymbol{\mathcal{A}}(t) = A_0 \left[ \boldsymbol{e}_1 \cos(\omega t) + \boldsymbol{e}_2 \sin(\omega t + \alpha) \right],
\label{eq:general_light}
\end{equation}
where $A_0$ is the amplitude, $\omega$ is the driving frequency, and the orthogonal unit vectors in the $xy$-plane are parameterized by the angle $\beta$:
\begin{equation}
\boldsymbol{e}_1 = (\cos\beta, \sin\beta, 0), \quad \boldsymbol{e}_2 = (-\sin\beta, \cos\beta, 0).
\end{equation}
Within this convention, different polarization states can be continuously tuned by varying $\alpha$ and $\beta$. Specifically, $\alpha = 0$ and $\pi$ describe circularly polarized light (CPL) with opposite helicities, whereas $\alpha = \pi/2$ and $3\pi/2$ represent linearly polarized light (LPL). 

In the following, we focus on the LPL case by setting:
\begin{equation}
\alpha = \frac{\pi}{2}, \quad \beta = 0.
\end{equation}
Under this selection, the unit vectors reduce to $\boldsymbol{e}_1 = (1, 0, 0)$ and $\boldsymbol{e}_2 = (0, 1, 0)$, yielding the diagonal LPL vector potential:
\begin{equation}
\boldsymbol{\mathcal{A}}(t) = \boldsymbol{A}_0 \cos(\omega t),
\label{eq:LPL_field}
\end{equation}
where $\boldsymbol{A}_0 = (A_0, A_0, 0)$ represents the polarization vector.

\subsection{Time-Dependent Hamiltonian and Peierls Substitution}
The pristine equilibrium tight-binding Hamiltonian for our 2D collinear AFM honeycomb model (with $t_{\text{NN}} = 1$, $m = 1$, and lattice constant $a = 1$) is formulated in momentum space as:
\begin{equation}
H_0(\boldsymbol{k}) = \sigma_0 \otimes \begin{pmatrix} 0 & \Delta_0(\boldsymbol{k}) \\ \Delta_0^*(\boldsymbol{k}) & 0 \end{pmatrix} + \sigma_z \otimes \tau_z,
\end{equation}
where the pristine hopping function is $\Delta_0(\boldsymbol{k}) = -\sum_{\ell=1}^3 e^{i \boldsymbol{k} \cdot \boldsymbol{\delta}_\ell}$. Incorporating the LPL field via the Peierls substitution $\boldsymbol{k} \rightarrow \boldsymbol{k} - \frac{e}{\hbar}\boldsymbol{\mathcal{A}}(t)$, the time-dependent hopping term becomes:
\begin{equation}
\Delta(\boldsymbol{k}, t) = -\sum_{\ell=1}^3 e^{i \boldsymbol{k} \cdot \boldsymbol{\delta}_\ell} e^{-i \frac{e}{\hbar} \boldsymbol{\mathcal{A}}(t) \cdot \boldsymbol{\delta}_\ell} = -\sum_{\ell=1}^3 e^{i \boldsymbol{k} \cdot \boldsymbol{\delta}_\ell} e^{-i \bar{A}_\ell \cos(\omega t)},
\end{equation}
where $\bar{A}_\ell = \frac{e}{\hbar} \boldsymbol{A}_0 \cdot \boldsymbol{\delta}_\ell$ is the dimensionless projection of LPL on the $\ell$-th bond. Using the Jacobi-Anger expansion, we decompose the time-dependent term into Fourier components:
\begin{equation}
e^{-i \bar{A}_\ell \cos(\omega t)} = \sum_{n=-\infty}^\infty (-i)^n J_n(\bar{A}_\ell) e^{i n \omega t}.
\end{equation}
Thus, the $n$-th Fourier component of the hopping function is:
\begin{equation}
\Delta_n(\boldsymbol{k}) = -\sum_{\ell=1}^3 e^{i \boldsymbol{k} \cdot \boldsymbol{\delta}_\ell} (-i)^n J_n(\bar{A}_\ell).
\end{equation}

\subsection{Proof of the Vanishing Second-Order Floquet Term under LPL}
In the high-frequency regime, the effective Floquet Hamiltonian is expanded as $H_{\text{eff}}(\boldsymbol{k}) = H_0(\boldsymbol{k}) + H^{(1)}(\boldsymbol{k}) + \mathcal{O}((\hbar\omega)^{-2})$, where the second-order effective term $H^{(1)}(\boldsymbol{k})$ (first-order in $1/(\hbar\omega)$) is determined by the commutator:
\begin{equation}
H^{(1)}(\boldsymbol{k}) = \frac{[H_{-1}(\boldsymbol{k}), H_1(\boldsymbol{k})]}{\hbar\omega}.
\end{equation}
The Fourier components $H_{\pm 1}(\boldsymbol{k})$ of the time-dependent Hamiltonian are given by:
\begin{equation}
H_1(\boldsymbol{k}) = \sigma_0 \otimes \begin{pmatrix} 0 & \Delta_1(\boldsymbol{k}) \\ \Delta_{-1}^*(\boldsymbol{k}) & 0 \end{pmatrix}, \quad
H_{-1}(\boldsymbol{k}) = \sigma_0 \otimes \begin{pmatrix} 0 & \Delta_{-1}(\boldsymbol{k}) \\ \Delta_{1}^*(\boldsymbol{k}) & 0 \end{pmatrix}.
\end{equation}
Evaluating the commutator explicitly yields:
\begin{equation}
[H_{-1}(\boldsymbol{k}), H_1(\boldsymbol{k})] = \left( |\Delta_{-1}(\boldsymbol{k})|^2 - |\Delta_1(\boldsymbol{k})|^2 \right) \sigma_0 \otimes \tau_z.
\end{equation}
For LPL, the Bessel function relations $J_{-1}(x) = -J_1(x)$ dictate the $n = 1$ and $n = -1$ Fourier components of the hopping term:
\begin{equation}
\Delta_1(\boldsymbol{k}) = i \sum_{\ell=1}^3 e^{i \boldsymbol{k} \cdot \boldsymbol{\delta}_\ell} J_1(\bar{A}_\ell), \quad
\Delta_{-1}(\boldsymbol{k}) = -i \sum_{\ell=1}^3 e^{i \boldsymbol{k} \cdot \boldsymbol{\delta}_\ell} J_{-1}(\bar{A}_\ell) = i \sum_{\ell=1}^3 e^{i \boldsymbol{k} \cdot \boldsymbol{\delta}_\ell} J_1(\bar{A}_\ell).
\end{equation}
Evidently, we have $\Delta_1(\boldsymbol{k}) = \Delta_{-1}(\boldsymbol{k})$, which directly implies:
\begin{equation}
|\Delta_{-1}(\boldsymbol{k})|^2 - |\Delta_1(\boldsymbol{k})|^2 = 0 \implies [H_{-1}(\boldsymbol{k}), H_1(\boldsymbol{k})] = 0.
\end{equation}
This mathematically proves that the second-order Floquet correction $H^{(1)}(\boldsymbol{k})$ vanishes identically under LPL. Thus, the effective Hamiltonian is purely determined by the time-averaged zeroth-order term:
\begin{equation}
H_{\text{eff}}^{\text{LPL}}(\boldsymbol{k}) \approx H_0(\boldsymbol{k}) = \sigma_0 \otimes \begin{pmatrix} 0 & \Delta_{\text{LPL}}(\boldsymbol{k}) \\ \Delta_{\text{LPL}}^*(\boldsymbol{k}) & 0 \end{pmatrix} + \sigma_z \otimes \tau_z,
\label{eq:H_LPL_eff}
\end{equation}
where $\Delta_{\text{LPL}}(\boldsymbol{k}) \equiv \Delta_0(\boldsymbol{k}) = -\sum_{\ell=1}^3 t_\ell e^{i \boldsymbol{k} \cdot \boldsymbol{\delta}_\ell}$ with $t_\ell = J_0(\bar{A}_\ell)$ denoting the light-renormalized anisotropic hopping.

\subsection{Valence Band Eigenstates of Decoupled Spin Sectors}
In the spinor basis, Eq.~\eqref{eq:H_LPL_eff} decouples into spin-up and spin-down blocks:
\begin{equation}
H_{\uparrow}(\boldsymbol{k}) = \begin{pmatrix} 1 & \Delta_{\text{LPL}}(\boldsymbol{k}) \\ \Delta_{\text{LPL}}^{*}(\boldsymbol{k}) & -1 \end{pmatrix}, \quad
H_{\downarrow}(\boldsymbol{k}) = \begin{pmatrix} -1 & \Delta_{\text{LPL}}(\boldsymbol{k}) \\ \Delta_{\text{LPL}}^{*}(\boldsymbol{k}) & 1 \end{pmatrix}.
\end{equation}
Writing $\Delta_{\text{LPL}}(\boldsymbol{k}) = |\Delta_{\text{LPL}}(\boldsymbol{k})| e^{i \theta(\boldsymbol{k})}$, the valence band eigenvalues are $E_- = -\lambda(\boldsymbol{k}) = -\sqrt{|\Delta_{\text{LPL}}(\boldsymbol{k})|^2 + 1}$. We define a parameterization angle $\vartheta(\boldsymbol{k})$ satisfying $\cos\vartheta(\boldsymbol{k}) = 1/\lambda(\boldsymbol{k})$ and $\sin\vartheta(\boldsymbol{k}) = |\Delta_{\text{LPL}}(\boldsymbol{k})|/\lambda(\boldsymbol{k})$. The normalized valence band eigenstates are derived as:
\begin{equation}
|u_{\text{val}, \uparrow}(\boldsymbol{k})\rangle = \begin{pmatrix} -\sin\left(\frac{\vartheta(\boldsymbol{k})}{2}\right) e^{i \theta(\boldsymbol{k})} \\ \cos\left(\frac{\vartheta(\boldsymbol{k})}{2}\right) \end{pmatrix}, \quad
|u_{\text{val}, \downarrow}(\boldsymbol{k})\rangle = \begin{pmatrix} -\cos\left(\frac{\vartheta(\boldsymbol{k})}{2}\right) e^{i \theta(\boldsymbol{k})} \\ \sin\left(\frac{\vartheta(\boldsymbol{k})}{2}\right) \end{pmatrix}.
\end{equation}

\subsection{Derivation of Spin-Resolved Polarizations}
Using the definition of the Berry connection $\boldsymbol{A}_{\text{val}, \sigma}(\boldsymbol{k}) = i \langle u_{\text{val}, \sigma} | \nabla_{\boldsymbol{k}} | u_{\text{val}, \sigma}\rangle$, we obtain the spin-resolved Berry connections:
\begin{equation}
\boldsymbol{A}_{\text{val}, \uparrow}(\boldsymbol{k}) = - \sin^2\left(\frac{\vartheta(\boldsymbol{k})}{2}\right) \nabla_{\boldsymbol{k}}\theta(\boldsymbol{k}) = - \frac{1}{2} \left( 1 - \frac{1}{\sqrt{|\Delta_{\text{LPL}}(\boldsymbol{k})|^2 + 1}} \right) \nabla_{\boldsymbol{k}}\theta(\boldsymbol{k}),
\end{equation}
\begin{equation}
\boldsymbol{A}_{\text{val}, \downarrow}(\boldsymbol{k}) = - \cos^2\left(\frac{\vartheta(\boldsymbol{k})}{2}\right) \nabla_{\boldsymbol{k}}\theta(\boldsymbol{k}) = - \frac{1}{2} \left( 1 + \frac{1}{\sqrt{|\Delta_{\text{LPL}}(\boldsymbol{k})|^2 + 1}} \right) \nabla_{\boldsymbol{k}}\theta(\boldsymbol{k}).
\end{equation}
Integrating these connections over the BZ yields the spin-resolved polarizations:
\begin{equation}
\mathbf{P}_{\uparrow} = \frac{e}{2(2\pi)^2} \int_{\mathrm{BZ}} d^2\boldsymbol{k} \left( 1 - \frac{1}{\sqrt{|\Delta_{\text{LPL}}(\boldsymbol{k})|^2 + 1}} \right) \nabla_{\boldsymbol{k}}\theta(\boldsymbol{k}),
\end{equation}
\begin{equation}
\mathbf{P}_{\downarrow} = \frac{e}{2(2\pi)^2} \int_{\mathrm{BZ}} d^2\boldsymbol{k} \left( 1 + \frac{1}{\sqrt{|\Delta_{\text{LPL}}(\boldsymbol{k})|^2 + 1}} \right) \nabla_{\boldsymbol{k}}\theta(\boldsymbol{k}).
\end{equation}
The sum of the spin-up and spin-down polarizations defines the total macroscopic polarization:
\begin{equation}
\mathbf{P}_{\text{total}} = \mathbf{P}_{\uparrow} + \mathbf{P}_{\downarrow} = \frac{e}{(2\pi)^2} \int_{\mathrm{BZ}} d^2\boldsymbol{k} \nabla_{\boldsymbol{k}}\theta(\boldsymbol{k}) = 0.
\end{equation}
Since $\mathbf{P}_{\text{total}} = 0$ is topologically guaranteed for this trivial insulator, we strictly obtain:
\begin{equation}
\mathbf{P}_{\uparrow} = -\mathbf{P}_{\downarrow} \neq 0,
\end{equation}
proving the emergence of a pure, light-induced compensated spin-antiferroelectric (Type-II AFE) state under LPL.

\bibliography{ref_sm}